\documentclass[onecolumn,superscriptaddress,amsmath,amssymb,aps,showkeys,floatfix]{revtex4-2}
\usepackage{amsmath,bm}
\usepackage{amssymb}
\usepackage{amsthm}
\usepackage{amsfonts}

\usepackage{mathtools}
\usepackage{rotating}
\usepackage{capt-of}
\usepackage{array}
\usepackage{blkarray, bigstrut}
\usepackage{enumitem}
\newlist{notes}{enumerate}{1}
\setlist[notes]{label=Note: ,leftmargin=*}
\usepackage{graphicx}
\usepackage{xcolor}
\usepackage{comment}
\usepackage{makecell}
\usepackage{hyperref}

\usepackage{capt-of}
\usepackage{float}
\makeatletter
\let\newfloat\newfloat@ltx
\makeatother

\usepackage{algorithm}
\usepackage{algpseudocode}
\usepackage{float}
\usepackage{enumitem}
\usepackage[section]{placeins}
\usepackage{bm}
\usepackage[normalem]{ulem}
\theoremstyle{definition}

\newcolumntype{P}[1]{>{\centering\arraybackslash}p{#1}}

\usepackage{multirow}
\usepackage{booktabs}
\usepackage{rotating,tabularx}
\usepackage{ulem}
\usepackage{url}
\usepackage{hhline}
\usepackage{siunitx}
\usepackage{dcolumn}
\renewcommand{\thetable}{\arabic{table}}

\makeatletter
\def\maketitle{
\@author@finish
\title@column\titleblock@produce
\suppressfloats[t]}
\makeatother
\newcommand{\cent}[1]{\centering\arraybackslash}

\renewcommand{\thesection}{\arabic{section}}

\makeatletter
\def\p@subsection{}
\makeatother
\makeatletter
\def\p@subsubsection{}
\makeatother
\usepackage{caption}
\usepackage{rotating}
\usepackage{booktabs}

\begin{document}
\preprint{APS}


\title{Bakry-\'Emery-Ricci curvature: An alternative network geometry measure in the expanding toolbox of graph Ricci curvatures}

\author{Madhumita Mondal}
\affiliation{The Institute of Mathematical Sciences (IMSc), Chennai 600113, India}
\affiliation{Homi Bhabha National Institute (HBNI), Mumbai 400094, India}
\affiliation{Max Planck Institute for Mathematics in the Sciences, 04103 Leipzig, Germany}

\author{Areejit Samal}
\thanks{To whom correspondence should be addressed:\\ asamal@imsc.res.in or muench@mis.mpg.de or jjost@mis.mpg.de}
\affiliation{The Institute of Mathematical Sciences (IMSc), Chennai 600113, India}
\affiliation{Homi Bhabha National Institute (HBNI), Mumbai 400094, India}

\author{Florentin M\"unch}
\thanks{To whom correspondence should be addressed:\\ asamal@imsc.res.in or muench@mis.mpg.de or jjost@mis.mpg.de}
\affiliation{Max Planck Institute for Mathematics in the Sciences, 04103 Leipzig, Germany}

\author{J\"urgen Jost}
\thanks{To whom correspondence should be addressed:\\ asamal@imsc.res.in or muench@mis.mpg.de or jjost@mis.mpg.de}
\affiliation{Max Planck Institute for Mathematics in the Sciences, 04103 Leipzig, Germany}
\affiliation{Center for Scalable Data Analytics and Artificial Intelligence, Leipzig University, 04109 Leipzig, Germany}
\affiliation{Santa Fe Institute for the Sciences of Complexity, Santa Fe, New Mexico 87501, USA}

\begin{abstract}
The characterization of complex networks with tools originating in geometry, for instance through the statistics of so-called Ricci curvatures, is a well established tool of network science. There exist various types of such Ricci curvatures, capturing different aspects of network geometry. In the present work, we investigate Bakry-\'Emery-Ricci curvature, a notion of discrete Ricci curvature that has been studied much in geometry, but so far has not been applied to networks. We explore on standard classes of artificial networks as well as on selected empirical ones to what the statistics of that curvature are similar to or different from that of other curvatures, how it is correlated to other important network measures, and what it tells us about the underlying network. We observe that most vertices typically have negative curvature. Random and small-world networks exhibit a narrow curvature distribution whereas other classes and most of the real-world networks possess a wide curvature distribution. When we compare Bakry-\'Emery-Ricci curvature with two other discrete notions of Ricci-curvature, Forman-Ricci and Ollivier-Ricci curvature for both model and real-world networks, we observe a high positive correlation between Bakry-\'Emery-Ricci and both Forman-Ricci and Ollivier-Ricci curvature, and in particular with the augmented version of Forman-Ricci curvature. Bakry-\'Emery-Ricci curvature also exhibits a high negative correlation with the vertex centrality measure and degree for most of the model and real-world networks. However, it does not correlate with the clustering coefficient. Also, we investigate the importance of vertices with highly negative curvature values to maintain communication in the network. Most of the networks are sensitive to the deletion of vertices with high degree and betweenness centrality. Additionally, for Forman-Ricci, Augmented Forman-Ricci, and Ollivier-Ricci curvature, we compare the robustness of the networks by comparing the sum of the incident edges and the minimum of the incident edges as vertex measures and find that the sum identifies vertices that are important for maintaining the connectivity of the network. The computational time for Bakry-\'Emery-Ricci curvature is shorter than that required for Ollivier-Ricci curvature but higher than for Augmented Forman-Ricci curvature. We therefore conclude that for empirical network analysis, the latter is the tool of choice.
\end{abstract} 

\maketitle 

\section{Introduction}
Network science examines the intricate structures and properties of networks formed by interconnected nodes or vertices and seeks general patterns or distinctive features across classes of networks \cite{networkscience2016}. Data from a wide range of fields are naturally modelled as networks, for instance biological systems such as Protein interaction networks \cite{rual2005towards}, social interactions such as Hamsterster or Facebook friendship networks, transportation systems such as the Chicago or Euro road networks \cite{eash1979equilibrium,vsubelj2011robust}, technological structures such as the US Power Grid, the Route views network \cite{leskovec2007graph} representing transformers or autonomous systems, and so forth. To investigate and characterize the properties of complex networks, geometry-inspired network measures have often been successfully applied \cite{sreejith2016forman,samal2018comparative, cushing2020bakry,elumalai2022graph,yadav2023discrete,fesser2023augmentations}.

In order to discover both universal patterns typically exhibited by empirical networks and to identify features that distinguish particular classes of networks, often methods are developed and used that have their origin in geometry. 
Among these methods, so-called discrete Ricci curvatures have been particularly useful. Originally, Ricci curvature was introduced in the smooth setting of Riemannian manifolds. It plays a fundamental role, for instance, in Einstein's theory of general relativity. In Riemannian geometry \cite{jost2008riemannian}, curvature quantifies how a geometric object deviates from a flat space. More recently, curvature notions have been generalized to the setting of metric spaces \cite{chow2003combinatorial, lott2009ricci, stone1976combinatorial,morgan2005manifolds, bonciocat2009mass,jost1997nonpositive,joharinad2019topology}. 

While in Riemannian geometry, there is a unique version of Ricci curvature, in the wider context of metric geometry, and also when metric geometry is narrowed down again to graphs, there are several notions of Ricci curvature. Each of them captures a different aspect of the original Riemannian notion. It is then natural to compare these notions and their statistics on model and empirical networks. Two prominent notions, Forman-Ricci and Ollivier-Ricci curvatures have already been systematically applied on networks \cite{sreejith2016forman,samal2018comparative,saucan2019discrete}. Here, we examine another notion, Bakry-\'Emery-Ricci curvature. That notion originated from a so-called Bochner identity \cite{gallot2004riemannian} that is used to control Laplacian eigenvalues in terms of lower Ricci bounds. In contrast to the other Ricci notions, it is evaluated on vertices instead of edges. It is then a natural question to which extent it captures similar or different aspects of networks than the other Ricci curvatures or other important network quantities. This is the topic of this paper.\\

Thus, in short, we shall theoretically and empirically compare three different types of graph Ricci curvatures. Due to conventions in geometry, the higher the effect measured, the more negative the curvature is.
\begin{itemize}
    \item Forman-Ricci curvature measures the spreading of information at the ends of an edge. The augmented version discounts triangles, in order to capture that spreading into different directions at the ends.
    \item Ollivier-Ricci curvature measures how far away the neighborhoods of the two ends of an edge are.
    \item Bakry-\'Emery-Ricci curvature measures the local cohesion of a network around a vertex, where more cohesion means more positive curvature.
\end{itemize}

In Section \ref{Theory}, we formally define and discuss three different notions of discrete Ricci curvatures, including of course Bakry-\'Emery-Ricci, but also the two prominent other versions due to Forman and Ollivier. Section \ref{Datasets} introduces the model and real-world networks for our subsequent analysis. Section \ref{Results} contains the main results about the correlations between the statistics of the different Ricci curvature and some other fundamental graph-theoretical quantities. The paper ends with a discussion of some directions for future works in Section \ref{Discusssion}.

\section{Theory} \label{Theory}

\subsection{Background} 
In Riemannian geometry, sectional curvature is a geometric property evaluated on two-dimensional tangent planes \cite{burago2022course,sher2001handbook,lua2005survey, saucan2005curvature}. It describes the convexity of the distance function between two geodesics. Ricci curvature is assigned to tangent directions, as the average sectional curvature across all tangent planes containing that direction. In the broader context of metric spaces, various notions of Ricci curvature have been proposed as generalizations of the Riemannian notion, and they capture different aspects of that notion. In graph theory Ricci curvature, whichever way defined, is usually evaluated on edges as the analogues of tangent directions. And in fact, it was argued in \cite{eidi2020edge}, that the analysis of graph properties via the statistics of local quantities should naturally concentrate on quantities that are evaluated on edges, as opposed to vertices, because it is the edges that capture the structure of a network.

Now, Bakry-\'Emery-Ricci curvature, which is the topic of this paper, is an exception from this general rule. It is a Ricci type curvature that is evaluated on vertices and not on edges. We should therefore provide some rationale for that. There is, for instance, a deep relation between Ricci curvature and eigenvalue bounds, both in Riemannian geometry and in graph theory, see for instance \cite{bauer2017geometric}. In particular, in the graph setting, a positive lower bound on Ricci yields a lower bound for the first non-vanishing and an upper bound for the largest eigenvalue of the graph Laplacian; for details see \cite{bauer2012ollivier}. But if we only care about lower bounds, then it makes sense to consider the minimum of the Ricci curvatures in all directions emanating from a vertex. This is a rationale for the Bakry-\'Emery-Ricci curvature which does precisely that, and in fact was developed around the scheme that produced the eigenvalue estimates in Riemannian geometry, the so-called Bochner identity. 

Actually, in Riemannian geometry, the Bochner identity involves other operators besides the Laplacian, in particular the Hessian \cite{reilly1977applications}. But those other operators cannot be directly defined on graphs, and so, the Bakry-\'Emery calculus circumvented that by introducing the $\Gamma$-products that can be expressed solely in terms of Laplace operators. The Bochner identity on a graph then looks more complicated and cumbersome than on a Riemannian manifold, but the important conceptual and practical gain is that in this way, the identity can be formulated solely in terms of the Laplacian of a graph. In fact, the formulation now is also such that it is not even necessary to refer to Ricci curvatures of edges, because the Laplacian flexibly interpolates between vertex and edge functions. Thus, a version of Ricci curvature is gained that only involves evaluating functions on vertices. 

Taking the minimum of Ricci curvatures is of course an operation that is different from taking sums. The latter operation in Riemannian geometry would yield scalar curvature, and scalar curvature is known to be an invariant that is much weaker than Ricci curvature. In contrast, lower bounds on Ricci curvatures have strong implications, not only for eigenvalues, but for instance also for the growth of the volumes of balls as function of their radii, see \cite{jost2008riemannian} and the references therein. Nevertheless, in graph theory, from other Ricci curvatures, Forman-Ricci and Ollivier-Ricci in this paper, we can naturally take sums at nodes, and then compare them with Bakry-\'Emery-Ricci.

 Recently, Bakry-\'Emery-Ricci curvature has received much interest in metric geometry and graph theory \cite{pouryahya2016bakry, pouryahya2017comparing,liu2018bakry,cushing2020bakry,cushing2022bakry,cushing2023bakry}. For instance, Pouryahya \textit{et al.} \cite{pouryahya2016bakry} find that cancer associated networks possess higher Bakry-\'Emery-Ricci curvature compared to non-cancer associated networks.

\begin{figure}
    \centering
    \includegraphics{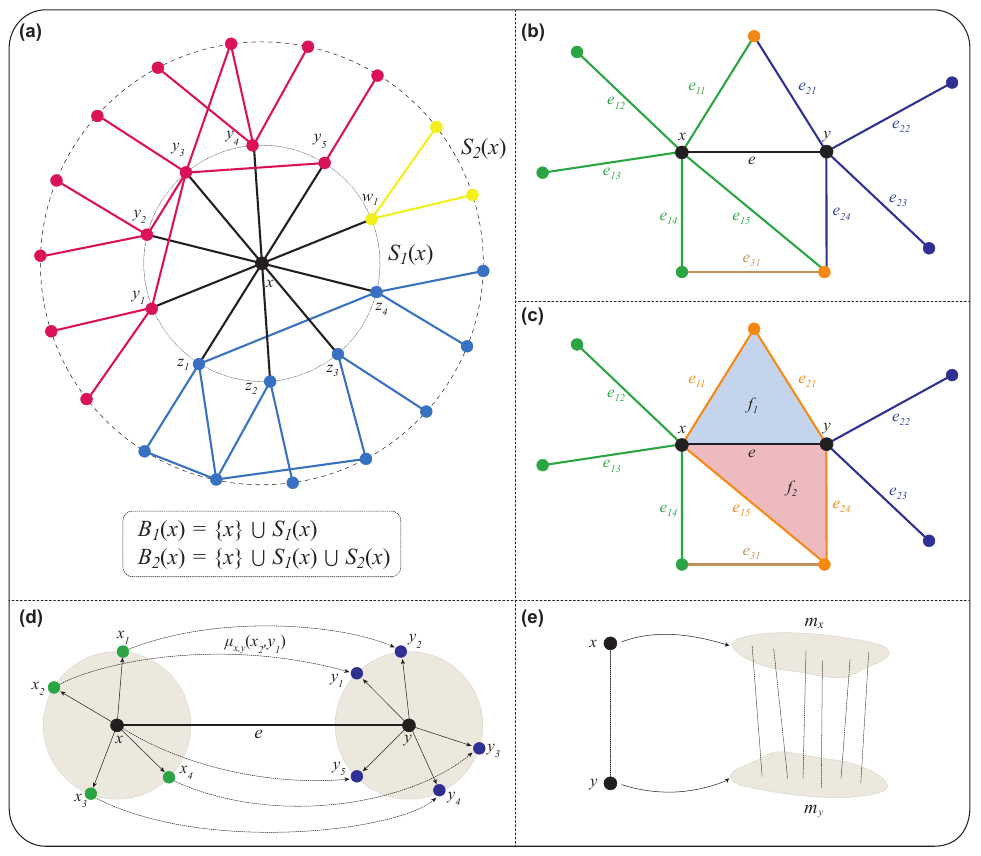}
    \caption{(\textbf{a}) Bakry-\'Emery-Ricci curvature of a vertex $x$. $S_1(x)$ and $S_2(x)$ contain the vertices present in the 1-sphere and the 2-sphere respectively. All the vertices included in the 2-ball $B_2(x)$ are taken into account for the computation of Bakry-\'Emery-Ricci curvature at vertex $x$. The colors indicate the connected components of the punctured ball $B_2(x)\backslash \{x\}$. The illustration is inspired by \cite{cushing2020bakry}. (\textbf{b,c}) Forman-Ricci curvature of an edge $e$ containing the vertices $x$ and $y$. Edges that share a child (a lower dimensional face) or a parent (a higher dimensional face) to the edge $e$ contribute to Forman-Ricci curvature. (b) Edges $e_{11}$, $e_{12}$, $e_{13}$, $e_{14}$, $e_{15}$ share a vertex $x$ with $e$ and edges $e_{21}$, $e_{22}$, $e_{23}$, $e_{24}$ share a vertex $y$ with $e$. Therefore all these edges are taken into account for the Forman-Ricci curvature of edge $e$. (c) For Augmented Forman-Ricci curvature of edge $e$, apart from all the edges mentioned above, the two-dimensional simplices $f_1$ and $f_2$ also contribute. (\textbf{d,e}) Illustration of Ollivier-Ricci curvature. (d) For a given vertex $x$, neighbors belong to the sphere $S_1(x)$ centered at $x$. Points on the sphere $S_1(y)$ centered at $x$ are transported to the points of the sphere centered at $y$. $\mu_{x, y}(x_2, y_1)$ is a probability measure from the vertex $x_2$ to the vertex $y_1$ through the edge $e$ with marginals equal to the measures $m_x$ and $m_y$ on $S_1(x)$ and $S_1(y)$, respectively. (e) An illustration of the measures $m_x$ and $m_y$. Points will be transported with average distance $(1-\kappa(x,y))d(x,y)$, where $\kappa(x,y)$ denotes the Ollivier-Ricci curvature between the nodes $x$ and $y$.}
    \label{fig:Schematic}
\end{figure}

\subsection{Bakry-\'Emery-Ricci curvature}

In the smooth setting of $n$-dimensional Riemannian manifolds, Bochner's identity \cite[Proposition 4.15]{gallot2004riemannian} provides a fundamental pointwise formula involving gradients, Laplacians, Hessians, and Ricci curvature. For all smooth functions $f \in C^\infty (M)$ where $(M,\langle.,.\rangle)$ is a Riemannian manifold of dimension $n$, Bochner's formula says
\begin{equation}
    \frac{1}{2}\Delta |\nabla f|^2(x) = |\text{Hess} ~ f|^2(x) + \langle \nabla \Delta f(x), \nabla f(x) \rangle + \text{Ric}(\nabla f(x),\nabla f(x)), \label{Bochneridentity}
\end{equation}
 where $\nabla$ and $\Delta$ stand for gradient and Laplacian respectively. Hess denotes the Hessian and $\mathrm{Ric}$ denotes the Ricci tensor. The Bakry-\'Emery-Ricci curvature notion is influenced by Bochner's identity and was introduced for smooth settings in \cite{bakry1985diffusions}. This curvature notion is determined purely in terms of the Laplace operator. In the discrete setting, this curvature notion has been studied extensively in \cite{schmuckenschlager1999curvature, hua2017stochastic, jost2014ollivier, lin2010ricci}.
Many theoretical results such as eigenvalue estimates \cite{lin2010ricci,bauer2017curvature}, diameter bounds \cite{liu2018bakry,liu2019distance,ambrosio2015bakry} and volume growth bounds
\cite{bauer2015li,horn2019volume,munch2019li} are based on gradient estimates for the heat equation \cite{lin2015equivalent,keller2018gradient,gong2017equivalent}.

For a finite simple graph $G=(V,E)$, where $V$ denotes the vertex set and $E$ denotes the edge sets, the (non-normalized) Laplacian $\Delta$ on any vertex $x\in V$ is defined by,
\begin{equation}
    \Delta f(x) = \sum_{y\sim x} (f(y)-f(x))
\end{equation}
where $f$ is a function $f:V\to \mathbb{R}$ and $y\sim x$ denotes that $y$ is a neighbour of $x$. Also one can consider the weighted graph Laplacian, however here for simplicity, we have considered the non-normalized graph Laplacian.

For any two functions $f,g:V\to\mathbb{R}$, the two operators $\Gamma$ and $\Gamma_2$ are defined as:
\begin{eqnarray}
    2 \Gamma(f,g) &=& \Delta (fg) - f\Delta g- g \Delta f \\
    2 \Gamma_2(f,g) &=& \Delta \Gamma(f,g)-\Gamma(f,\Delta g)-\Gamma(g,\Delta f).
\end{eqnarray}
Here, $\Gamma(f,f)$ is analogous to $|\nabla f|^2(x)$ and $\Gamma_2(f,f)$ is analogous to $\frac{1}{2}\Delta |\nabla f|^2(x) - \langle \nabla \Delta f(x), \nabla f(x) \rangle$ with Bochner identity (see Eq. (\eqref{Bochneridentity})). The Bakry-\'Emery-Ricci curvature on a vertex $x \in V$ (non-isolated) is defined as the supremum of all values of $K \in \mathbb{R}$ that satisfy the curvature-dimension inequality $CD(K,n)$ for a fixed dimension $n \in (0,\infty]$ and for all functions $f:V\to \mathbb{R}$,
\begin{equation}
     \Gamma_2(f)(x) \geqslant \frac{1}{n}(\nabla f(x))^2 + K \Gamma(f)(x)
\end{equation}
where $\Gamma(f) = \Gamma(f,f)$, $\Gamma_2(f) = \Gamma_2(f,f)$. Therefore, for a finite simple graph $G=(V,E)$, the Bakry-\'Emery-Ricci curvature on a vertex $x \in V$ can be found by taking $n \to \infty$ and solving the following semidefinite programming problem,
\begin{eqnarray}
    \text{maximize}\hspace{1cm} && K \\
    \text{subject to}\hspace{0.95cm} && \Gamma_2(f)(x) \geqslant K \Gamma(f)(x).
\end{eqnarray}
Now, for a better understanding of the local matrices $\Delta f(x)$, $\Gamma(f)(x)$ and $\Gamma_2(f)(x)$, we define the $r$-ball $\forall r \in \mathbb{N}$ centered at $x$ as, $B_r(x) = \{y \in V : dist(x,y) \le r\}$, and the $r$-sphere centered at $x$ as, $S_r(x) = \{y \in V : dist(x,y) = r\}$, where $dist(x,y)$ is the shortest path length between the vertices $x$ and $y$. Then, we can write $B_2(x) = \{x\} \cup S_1(x) \cup S_2(x)$.

However, the size of the matrices $\Delta f(x)$, $\Gamma(f)(x)$ and $\Gamma_2(f)(x)$ are $1 \times |{V}|$, $|{V}| \times |{V}|$ and $|{V}| \times |{V}|$ respectively, the matrices are typically sparse. The non-trivial blocks of $\Delta f(x)$ is of size $1 \times |{B_1(x)}|$ = $1 \times (\deg(x)+1)$, where $\deg(x)$ is the degree of the vertex $x$. The non-trivial blocks of $\Gamma(f)(x)$ and $\Gamma_2(f)(x)$ are of size $|B_1(x)| \times |{B_1(x)}|$ and $|{B_2(x)}| \times |{B_2(x)}|$ respectively. Specifically, the Bakry-\'Emery-Ricci curvature value at vertex $x$ only depends on the graph structure of $B_2(x)$. For more details about the local structure of $\Gamma(f)$ and $\Gamma_2(f)$, readers are recommended to \cite[Section 2]{cushing2020bakry}. If the punctured 2-ball defined as $\stackrel{\circ}{B}_2(x) = B_2(x) - \{x\}$ of a vertex $x$, contains more than one connected component then in most cases we get a negative curvature value at vertex $x$ \cite{cushing2020bakry}. Fig. \ref{fig:Schematic}(a) illustrates the $S_1(x)$, $S_2(x)$, $B_1(x)$ and $B_2(x)$. The punctured 2-ball i.e. the 2-ball after removing the vertex $x$ separated the network into three different connected components, which leads to a negative Bakry-\'Emery-Ricci curvature of the vertex $x$ (= -4.545). The Bakry-\'Emery-Ricci curvature has been calculated using a Mathematica code which is a modified version of the code available at: \url{https://sites.google.com/view/florentin-muench/graph-curvature?authuser=0}. Indeed, Bakry-\'Emery-Ricci curvature does not just control the local topology of the network, but also determines the homology of a graph. That is, if the curvature is positive or mostly positive, then there can not be any large holes in the graph \cite{kempton2022homology,munch2020spectrally}.

\subsection{Forman-Ricci curvature}

Forman's discretization of classical Ricci curvature was defined for weighted CW cell complexes \cite{whitehead1949combinatorial} which is a particular class of geometric objects \cite{forman2003bochner}. An important point to note is that Forman's definition of classical Ricci curvature in the context of weighed CW cell complexes develops a proper explanation of the Bochner-Weitzenb\"ock formula \cite{jost2008riemannian}, which establishes a relationship between curvature and classical (Riemannian) Laplace operator. Like other curvature notions, Forman-Ricci curvature captures the contraction rates under the diffusion \cite[Theorem 1.1]{jost2021characterizations}. 

From Riemannian geometry, the Bochner-Weitzenb\"ock formula states that $H=B+\mathrm{Ric}$ where $H$ is the Hodge Laplacian, $B$ the Bochner Laplacian and $\mathrm{Ric}$ the Ricci curvature. The Hodge Laplacian is an abstract operator which can be defined in a very general setting such as chain complexes. The Bochner Laplacian satisfies a certain maximum principle, i.e., the semigroup of the Bochner-Laplacian is contractive.
In the discrete setting, both Bochner and Hodge Laplacian act on edges. The maximum principle for the Bochner Laplacian translates to diagonal dominance, and the difference between Hodge and Bochner Laplacian is required to be a diagonal matrix, which will define the Forman-Ricci curvature.
More concretely, given a coboundary operator $\delta$ on a cell complex and the corresponding Hodge Laplacian matrix $H=\delta \delta^* + \delta^*\delta$, the Forman-Ricci curvature of an edge $e$ is defined as
\begin{equation}
    F(e) = H_{e,e} - \sum_{e' \neq e} \frac{\ell(e')}{\ell(e)}|H_{e,e'}| \label{FRCHodgeLaplacian}
\end{equation}
where $\ell(e)$ denotes the length of the edge and can be typically set to be one, see \cite[Section~7]{jost2021characterizations}. 
More explicitly, for a regular cell complex with cell weights $m$, and an edge $e$ with endpoints $x,y$, 
\begin{equation}
   F(e) = \frac{m(e)}{m(x)} + \frac{m(e)}{m(y)} + \sum_{f>e} \frac{m(f)}{m(e)}- \sum_{e' \neq e} \frac{\ell(e')}{\ell(e)} \left| \sum_{z<e,e'} \frac{m(e')}{m(v')} 
 - \sum_{f>e,e'} \frac{m(f)}{m(e)}\right| \label{FRCweighted} 
\end{equation}
where $z$ is the common vertex of the edges $e'$ and $e$, and $f$ is a face containing the edge $e$, or both edges $e$ and $e'$ respectively. To translate to Forman's original formula for cell complexes with cell weights $w$, one chooses $m=1/w$ and $\ell = \sqrt{w}$, and multiplies the curvature by $w$ (see Fig. \ref{fig:Schematic}(b)). We remark that in case of unit weights, the definition here coincides with Forman's definition. The reason why we do not use Forman's original definition for the weighted case is that our definition has better compatibility with Ollivier-Ricci curvature, and thus gives desired theoretical implications such as eigenvalue estimates and diameter bounds \cite{jost2021characterizations}.

For more details on the general concept of discrete Ricci curvatures, we refer the interested reader to Forman's original work \cite{forman2003bochner} and our previous contribution wherein we were the first to adapt the notion of Forman-Ricci curvature to undirected, directed and higher-order networks
\cite{sreejith2016forman,saucan2019discrete,saucan2018discrete,samal2018comparative}. 

In terms of combinatorial graphs, the weights of all the edges and vertices are considered as $1$, i.e., $m \equiv 1 \equiv \ell$, then Eq. (\eqref{FRCweighted}) simplifies to a more straightforward form,
\begin{equation}
    F(e) = 4 - \sum_{v\sim e}\deg(v) \label{FRCunweighted}
\end{equation}
 where $v\sim e$ denotes the endpoint vertices of edge $e$. Intuitively, Forman-Ricci curvature measures how widely information spreads from the endpoints of an edge in a network. An edge with a highly negative curvature value indicates that a significant amount of information disperses from its endpoints. Mainly, Forman-Ricci curvature is defined on the edges yet it is also defined on a vertex by taking the sum of its incident edges \cite{samal2018comparative,sreejith2017systematic}. However, in this contribution, we define vertex curvature value as the minimum of its incident edge curvature values.

 \textbf{Augmented Forman-Ricci curvature:} 
  We have also considered the modified version of Forman-Ricci curvature namely Augmented Forman-Ricci curvature \cite{samal2018comparative}, which takes into account the clique complex, the simplicial complex containing all complete subgraphs as simplices. Particularly, the Augmented Forman-Ricci curvature also takes triangles into account. Augmented Forman-Ricci curvature inspects only the triangles while neglecting cycles with lengths greater than three (see Fig. \ref{fig:Schematic}(c)).

For an unweighted network, one has $m(v) = m(e)=\ell(e) = m(f) = 1$ for all $f\in F(G)$, and $e\in E(G)$, and $v\in V(G)$, where $F(G)$, $E(G)$ and $V(G)$ denotes the set of triangular faces, edges, and vertices in the graph $G$ respectively. For unweighted networks, we have a straightforward relationship between Forman-Ricci and Augmented Forman-Ricci curvature \cite{samal2018comparative} given by,
\begin{equation}
    F^{\#}(e) = F(e) + 3m
\end{equation}
 where $m$ represents the number of triangles in the network that contain the edge $e$. Similar to Forman-Ricci curvature, we define vertex curvature value as the minimum of its incident edge curvature values on that vertex.

\subsection{Ollivier-Ricci curvature}

The discretization of Ricci curvature by Ollivier \cite{ollivier2007ricci, ollivier2009ricci, ollivier2010survey} captures a different aspect of the classical (smooth) notion. In this approach, if $B_x$ and $B_y$ are two different small balls in an $n$-dimensional Riemannian manifold with center $x$ and $y$ respectively and with radius $\varepsilon$, then the average distance between the points on $B_x$ and their corresponding points on $B_y$ when we arrange the correspondence by optimal transport, to be defined \cite{vaserstein1969markov} below, is:
\begin{equation}
    \delta\left(1-\frac{\varepsilon^2}{2(n+2)} \operatorname{Ric}(v) + O(\varepsilon^3 + \varepsilon^2\delta)\right)
\end{equation}
where $\delta$ and $v$ denote the distance ($d(x,y)$) and tangent vector between $x$ and $y$ respectively. Furthermore $\delta, \varepsilon \to 0$. Therefore, positive (negative) curvature means that the balls are closer (more distant) to each other compared to the distance between their centers. By definition, Ollivier-Ricci curvature between the points $x$ and $y$ is given by
\begin{eqnarray}
    \kappa(x,y) &=& \lim \limits_{\varepsilon \to 0}\frac{1}{\varepsilon}\kappa_\varepsilon(x, y)\\
     \kappa_\varepsilon(x, y) &=& 1-\frac{W_1\left(m_x^\varepsilon, m_y^\varepsilon\right)}{d(x, y)} 
\end{eqnarray}
where $W_1(m_x^\varepsilon, m_y^\varepsilon)$ denote the Wasserstein transportation metric between $m_x^\varepsilon$ and $m_y^\varepsilon$, the measure of the ball around $x$ and $y$ respectively. Between the two probability measures $m_x^\varepsilon$ and $m_y^\varepsilon$, the Wasserstein distance or the transportation distance $W_1(m_x^\varepsilon, m_y^\varepsilon)$ is defined as,
\begin{equation}
    W_1\left(m_x^\varepsilon, m_y^\varepsilon\right)=\inf _{\mu_{x, y} \in \prod\left(m_x^\varepsilon, m_y^\varepsilon\right)} \sum_{\left(x^{\prime}, y^{\prime}\right) \in V \times V} d\left(x^{\prime}, y^{\prime}\right) \mu_{x, y}\left(x^{\prime}, y^{\prime}\right)
\end{equation}
where $\prod\left(m_x^\varepsilon, m_y^\varepsilon\right)$ is the set of probability measures $\mu_{x, y}$ satisfying,
\begin{equation}
    \sum_{y^{\prime} \in V} \mu_{x, y}\left(x^{\prime}, y^{\prime}\right)=m_x^\varepsilon\left(x^{\prime}\right),~~ \sum_{x^{\prime} \in V} \mu_{x, y}\left(x^{\prime}, y^{\prime}\right)=m_y^\varepsilon\left(y^{\prime}\right). \label{ProbabilityMeasures}
\end{equation}
The probability measures $m_x^{\varepsilon}(z)$ for sufficiently small $\varepsilon$, can be seen as
\begin{equation}
  m_x^{\varepsilon}(z) = \begin{cases}
  1-\varepsilon \deg(x) & \text{if } z = x \\
  \varepsilon & \text{if } z \sim x 
\end{cases}
\end{equation}
where $z\sim x$ denotes that $z$ is a neighbor of $x$. The transportation problem starts with the probability measure $m_x^{\varepsilon}$ at vertex $x$ and ends at vertex $y$ with the probability measure $m_y^{\varepsilon}$ after considering all possible transportation options, that satisfy the Eq. (\eqref{ProbabilityMeasures}). $W_1\left(m_x^\varepsilon, m_y^\varepsilon\right)$ represent the minimal cost of transportation of mass from $m_x^{\varepsilon}$ to $m_y^{\varepsilon}$. Here, we have accounted for the non-normalized probability measures as in the case of Bakry-\'Emery-Ricci curvature we have considered the non-normalized Laplacian. 
The Ollivier-Ricci curvature can also be expressed purely in terms of the graph Laplacian $\Delta$, namely,
\[
\kappa(x,y) = \inf_{\substack{f \in Lip(1)\\
f(y)-f(x)=d(x,y)}} \frac{\Delta f(x)-\Delta f(y)}{d(x,y)},
\]
see \cite{munch2019ollivier}. Considering different weights for the Laplacian corresponds to different choices of $m_x^\varepsilon$.
The normalized Laplacian or probability measure, i.e., dividing the probability measures by the degree was examined in some of our previous works \cite{samal2018comparative,elumalai2022graph,yadav2023discrete}.

For a combinatorial graph, the Ollivier-Ricci curvature of a given edge $(x,y)$ is determined as follows \cite{munch2019ollivier},
\begin{itemize}
    \item Initial curvature of the edge $(x,y)$ is 2.
    \item For each triangle involving the edge $(x,y)$ increase the curvature value by 1.
    \item Presence of a 4-cycle does not affect the curvature value of the edge $(x,y)$.
    \item Presence of a 5-cycle decreases the curvature value of the edge $(x,y)$ by 1.
    \item Each additional neighbor of the vertices $x$ and $y$ decreases the curvature value of the edge $(x,y)$ by 1.
\end{itemize}
Note that, the presence of a triangle dominates the presence of 4-cycles and the presence of 4-cycles dominates the presence of 5-cycles.

Ollivier-Ricci curvature is also defined on an edge of a network. However, we can utilize this discrete notion also as a vertex measure. Like Forman-Ricci and Augmented Forman-Ricci, Ollivier-Ricci curvature at a vertex $x$ is defined by the minimum of its incident edge curvature values. Recently it was shown that the Ollivier-Ricci curvature of a graph coincides with the Forman-Ricci curvature of a 2-dimensional cell complex having the graph as one skeleton \cite[Theorem 1.2]{jost2021characterizations}.


\section{Dataset of model and real-world networks} 
\label{Datasets}

We have considered both model and real-world undirected and unweighted networks for this analysis of Bakry-\'Emery-Ricci curvature and its comparison with other graph Ricci curvatures. The model networks considered are: 
\begin{itemize}
    \item \textbf{Erd\"os-R\'enyi (ER) model} \cite{engel2004large}: This model generates a collection of random graphs $G(n,p)$, where $n$ denotes the total number of vertices and $p$ denotes the probability of each edge to be included in the network. This probability is independent of the inclusion of other edges.
    \item \textbf{Watts-Strogatz (WS) model} \cite{watts1998collective}: This model generates random graphs $G(n,k,p)$ with small-world property, i.e., graphs exhibiting both high clustering coefficient and short average path length. This model starts from a regular ring lattice with $n$ vertices where each vertex is connected with its $k$ nearest neighbors. Then one endpoint of each edge is rewired with a probability $p$ to another vertex of the network. The new vertex is chosen uniformly at random while avoiding link duplication and self-loops.
    \item \textbf{Barab\'asi-Albert (BA) model} \cite{barabasi1999emergence}: This model generates networks $G(n,m)$ with scale-free property, i.e., the generated graphs exhibit a power-law degree distribution. Initially, the starting graph contains $m_0$ ($m_0 \ge m$) vertices. Thereafter, at each step, one vertex is introduced with degree $m$ which connects to the existing vertices based on preferential attachment scheme, i.e., the probability of an existing vertex to receive a new connection is proportional to its degree. 
    \item \textbf{Hyperbolic Graph Generator (HGG) model} \cite{krioukov2010hyperbolic}: This model generates random hyperbolic graphs $G(n,k)$ with power-law degree distribution and non-vanishing clustering. On a hyperbolic disk, $n$ vertices are placed randomly, and subsequently, two vertices are connected with a probability depending on their hyperbolic distance. The input parameters to generate a HGG model are number of vertices $n$, targeted average degree $k$, desired power-law degree distribution exponent $\gamma$, and temperature $T$. For this analysis, the hyperbolic random geometric graphs were generated by considering the default input parameter values of $\gamma$ = 2 and $T$ = 0.
\end{itemize}
The details of the analyzed model networks like the number of vertices, number of edges, fraction of vertices in the largest connected component, average degree, edge density, and average shortest path length are listed in SI Table \ref{Model_details}. Further, while analyzing the Forman-Ricci, Augmented Forman-Ricci, Ollivier-Ricci, and Bakry-\'Emery-Ricci curvatures, we have used the largest connected component of the considered networks.

Along with the model networks, we have analyzed undirected and unweighted real-world networks. Specifically, we have considered four infrastructure networks: Chicago road \cite{eash1979equilibrium}, US Power Grid \cite{leskovec2007graph}, Euro road \cite{vsubelj2011robust}, and Contiguous US States \cite{knuth2006art}; four biological networks: Human protein interactions \cite{rual2005towards}, Reactome \cite{joshi2005reactome}, PDZ-domain \cite{beuming2005pdzbase}, and Yeast protein interactions \cite{jeong2001lethality}; four social networks: Facebook \cite{leskovec2012learning}, Jazz musicians \cite{gleiser2003community}, Hamsterster friendships, and Zachary karate club \cite{zachary1977information}; one communication network Email communication \cite{guimera2003self}; one linguistic network Adjective-Noun adjacency \cite{newman2006finding}; one co-authorship network: Astrophysics co-authorship \cite{leskovec2007graph}; one online contact network: Pretty Good Privacy (PGP) \cite{boguna2004models}; one computer network: Route views \cite{leskovec2007graph}; one interaction network: Football \cite{girvan2002community}; and two species networks: Dolphin \cite{lusseau2003bottlenose} and Zebra \cite{sundaresan2007network}. The basic characteristics of these real-world networks are as follows:
\begin{itemize}
    \item \textbf{Chicago road network} \cite{eash1979equilibrium} contains 1467 vertices and 1298 edges, wherein the vertices correspond to transportation zones within the Chicago region.
    \item \textbf{US Power Grid} \cite{leskovec2007graph} contains 4941 vertices corresponding to the generators or transformers or substations in the western states of the USA and 6594 edges corresponding to power supply lines.
    \item \textbf{Euro road network} \cite{vsubelj2011robust} contains 1174 vertices and 1417 edges. The vertices correspond to European cities and the edges correspond to road links in the international E-road network.
    \item \textbf{Contiguous US States network} \cite{knuth2006art} contains 49 vertices representing 48 contiguous states of USA and the District of Columbia (excluding Alaska and Hawaii due to their lack of land connection with the other states) and 107 edges representing the presence of land border between the states.
    \item \textbf{Human protein interactions network} \cite{rual2005towards} contains 2783 vertices representing proteins and 6007 edges representing the interactions between human proteins as recorded in a prior version of the proteome-scale map of binary protein interactions in humans.
    \item \textbf{Reactome network} \cite{joshi2005reactome} contains 6327 vertices representing proteins in humans and 147547 edges representing interaction among proteins sourced from the Reactome project, an open online database of pathways.
    \item \textbf{PDZ-domain interaction network} \cite{beuming2005pdzbase} contains 212 vertices corresponding to proteins and 242 edges representing PDZ-domain-mediated interactions between two proteins.
    \item \textbf{Yeast protein interactions network} \cite{jeong2001lethality} contains 1846 vertices corresponding to proteins in yeast \textit{Saccharomyces cerevisiae} and 2203 edges corresponding to interaction between two proteins.
    \item \textbf{Facebook network} \cite{leskovec2012learning} contains 2888 vertices representing the Facebook users and 2981 edges indicating friendship among the Facebook users.
    \item \textbf{Jazz musicians network} \cite{gleiser2003community} contains 198 vertices corresponding to Jazz musicians and 2742 edges representing collaboration between musicians.
    \item \textbf{Hamsterster friendships network} contains 1858 vertices representing the users of hamsterster.com and 12534 edges representing friendships between the users. 
    \item \textbf{Zachary karate club network} \cite{zachary1977information} contains 34 vertices corresponding to members of a university karate club by Wayne Zachary and 78 edges representing ties between members.
    \item \textbf{Email communication network} \cite{guimera2003self} contains 1133 vertices representing the users at the University Rovira i Virgili in Tarragona in South Catalonia in Spain and 5451 edges representing direct communication between users.
    \item \textbf{Adjective-Noun adjacency network} \cite{newman2006finding} contains 112 vertices corresponding to nouns or adjectives and 425 edges representing their co-occurrence in adjacent positions within the novel `David Copperfield' authored by Charles Dickens.
    \item \textbf{Astrophysics co-authorship network} \cite{leskovec2007graph} contains 18771 vertices corresponding to authors of scientific papers from arXiv's Astrophysics section and the 198050 edges representing co-authored publications by two authors.
    \item \textbf{Pretty Good Privacy (PGP) network} \cite{boguna2004models} contains 10680 vertices corresponding to users of the Pretty Good Privacy (PGP) algorithm and 24316 edges representing the interactions between the users.
    \item \textbf{Route views network} \cite{leskovec2007graph} contains 6474 vertices representing autonomous systems and 13895 edges denoting communication between the autonomous systems.
    \item \textbf{Football network} \cite{girvan2002community} contains 115 vertices representing American football teams of Division IA colleges during the regular season of the year 2000 and 613 edges corresponding to games between them.
    \item \textbf{Dolphin network} \cite{lusseau2003bottlenose} contains 62 vertices representing bottlenose dolphins within a dolphin community residing off Doubtful Sound, a fjord in New Zealand, and 159 edges representing frequent associations among Dolphins. The observations were made between 1994 and 2001.
    \item \textbf{Zebra network} \cite{sundaresan2007network} contains 27 vertices representing Gr\'evy's zebras in Kenya and 111 edges representing an interaction between them during the study.
\end{itemize}
Most of the undirected real-world networks mentioned above were obtained from the KONECT database \cite{kunegis2013konect}. The details of the analyzed real-world networks like the number of vertices, number of edges, fraction of vertices in the largest connected component, average degree, edge density, and average shortest path length are listed in SI Table \ref{Real_details}. Since the considered networks are unweighted, we have considered the weights of the vertices, edges, and two-dimensional simplicial complexes as 1 during the computation of curvatures.



\section{Results} \label{Results}
\subsection{Distribution of Bakry-\'Emery-Ricci curvature in model and real-world networks}

\begin{figure}
    \centering
    \includegraphics{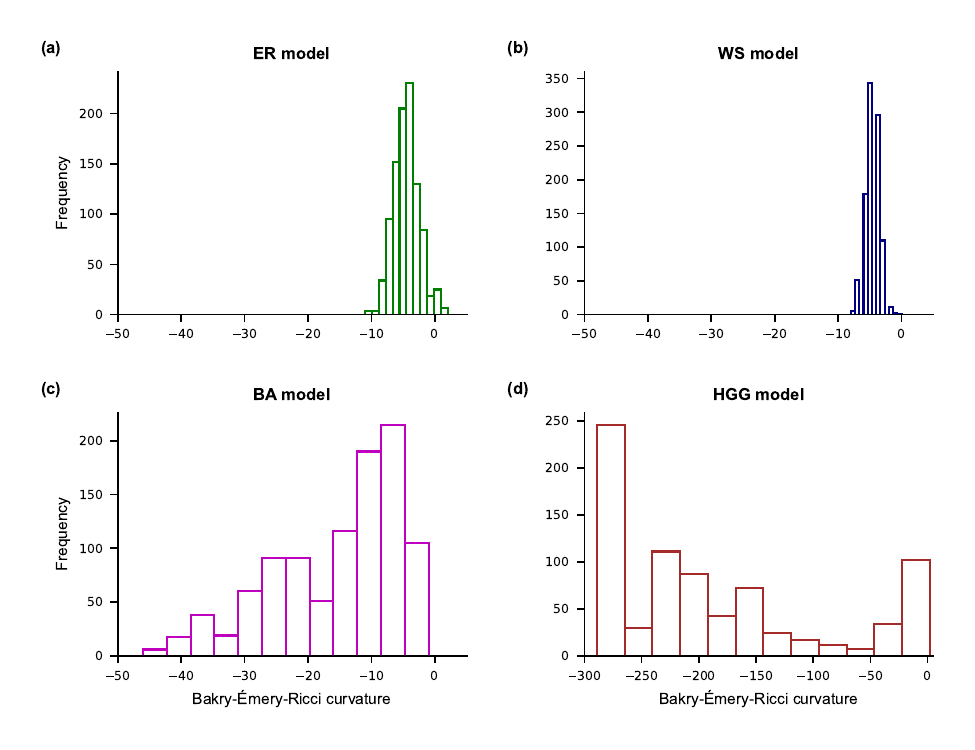}
    \caption{Distribution of Bakry-\'Emery-Ricci curvature for different model networks with the number of vertices, $n$ = 1000. (a) ER model with $p$ = 0.005, (b) WS model with $k$ = 6, $p$ = 0.5, (c) BA model with $m$ = 3 and (d) HGG model with $k$ = 3, $\gamma$ = 2, $T$ = 0.}
    \label{Distribution_Model}
\end{figure}

Distributions of Bakry-\'Emery-Ricci curvature for the model networks are illustrated in Fig. \ref{Distribution_Model}. The curvature value of each vertex is mostly negative for different types of model networks. The distributions of vertex curvature for the random networks generated by the ER model and the small-world networks generated by the WS model are narrower while the distributions of vertex curvature for the scale-free network generated by the BA model and the random hyperbolic graph generated by the HGG model are wider. Both the ER and WS models follow a Poisson degree distribution. In contrast, BA and HGG follow a power-law degree distribution and in general, curvature values decrease for high-degree vertices. Vertex curvature values decrease by nearly -10, particularly for ER and WS models. However, the values plummet close to -46 and -289 for BA and HGG models respectively. Therefore, there is a clear difference in the vertex curvature distribution for different types of model networks.

\begin{figure}
    \centering
    \includegraphics{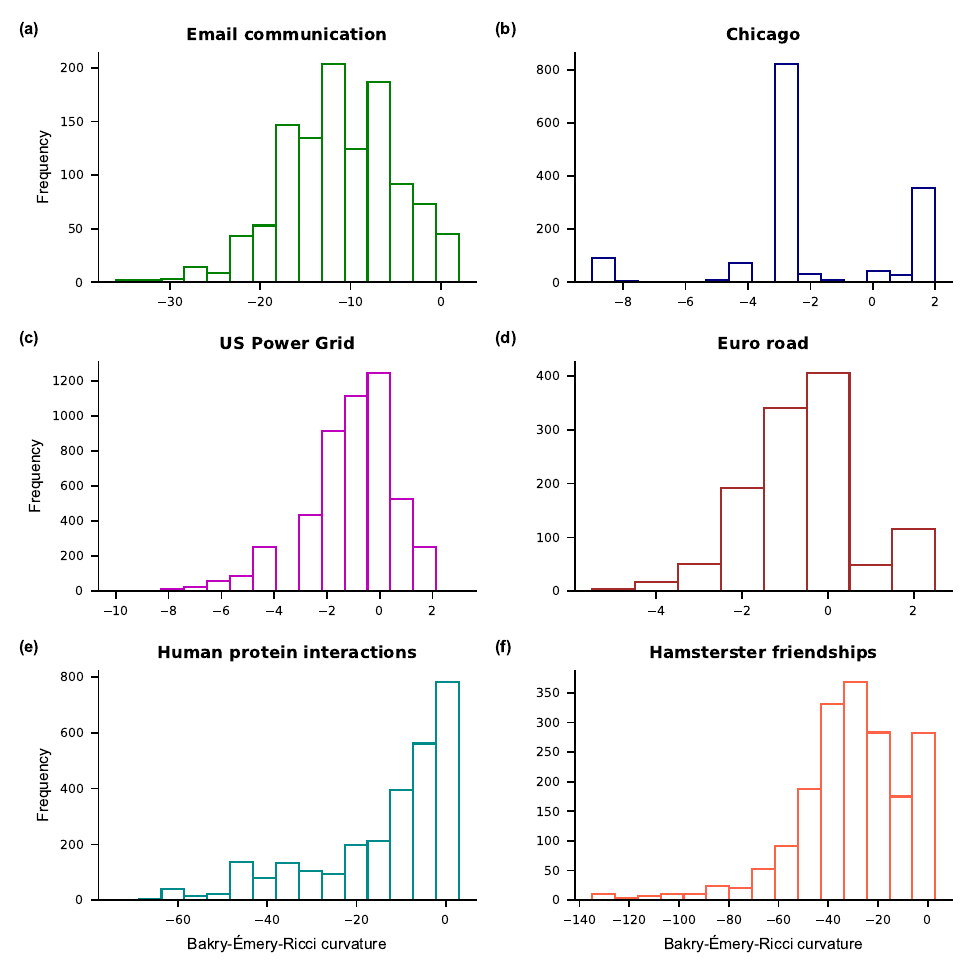}
    \caption{Distribution of Bakry-\'Emery-Ricci curvature for different real-world networks (a) Email communication, (b) Chicago road network, (c) US Power Grid, (d) Euro road network, (e) Human protein interactions, and (f) Hamsterster friendships networks.}
    \label{Distribution_real}
\end{figure}

In Fig. \ref{Distribution_real}, the distributions of Bakry-\'Emery-Ricci curvature are shown for six different real-world networks. Analogous to the Bakry-\'Emery-Ricci curvature distribution of the model networks, real-world networks also contain mostly negative curvature values on their vertices. In particular, the Email communication, Human protein interaction  and Hamsterster friendships networks contain some vertices with highly negative curvature values. Bakry-\'Emery-Ricci curvature distributions corresponding to the other real-world networks have been shown in SI Fig. \ref{Distribution_realothers}.

\subsection{Correlation between Bakry-\'Emery-Ricci curvature and other measures}
\subsubsection{Correlation with Forman-Ricci, Augmented Forman-Ricci, and Ollivier-Ricci curvature}

\begin{figure}
    \centering
    \includegraphics{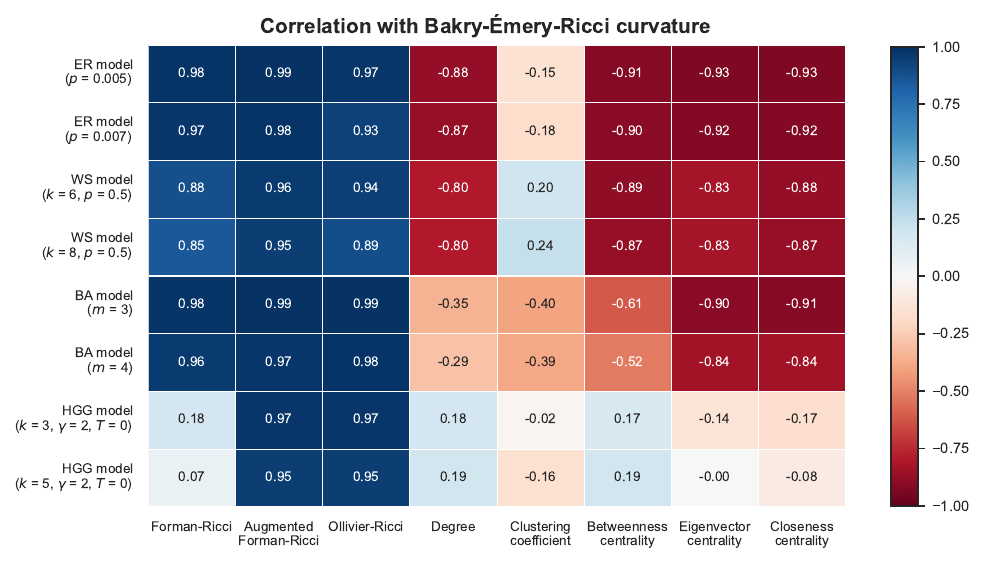}
    \caption{Spearman correlation (rounded to two decimal places) of Bakry-\'Emery-Ricci curvature with Forman-Ricci curvature, Augmented Forman-Ricci curvature, Ollivier-Ricci curvature, degree, clustering coefficient, betweenness centrality, eigenvector centrality, and closeness centrality in model networks. The columns correspond to the eight different vertex measures. The rows correspond to different model networks.}
    \label{heatmap_Model}
\end{figure}

\begin{figure}
    \centering
    \includegraphics{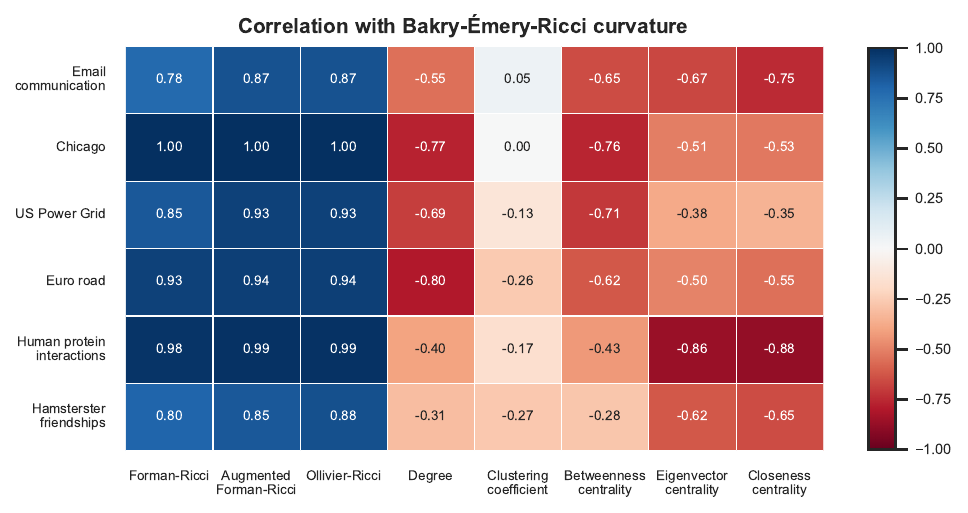}
    \caption{Spearman correlation (rounded to two decimal places) of Bakry-\'Emery-Ricci curvature with Forman-Ricci curvature, Augmented Forman-Ricci curvature, Ollivier-Ricci curvature, degree, clustering coefficient, betweenness centrality, eigenvector centrality, and closeness centrality in six real-world networks. The columns correspond to the eight different vertex measures. The rows correspond to six different real-world networks.}
    \label{heatmap_Real}
\end{figure}

We have compared the Bakry-\'Emery-Ricci curvature with Forman-Ricci, Augmented Forman-Ricci, and Ollivier-Ricci curvature. For model networks, we have generated 100 networks for each parameter value and then considered the mean of the correlations. Further, we have performed the correlation considering the largest connected components of the networks. We have observed a very high correlation between the Bakry-\'Emery-Ricci curvature and other vertex curvature measures in the ER, WS, and BA model networks. However, for the HGG model, Bakry-\'Emery-Ricci curvature is not correlated with Forman-Ricci curvature, as the HGG networks contain many triangles and triangles play an important role for Bakry-\'Emery-Ricci curvature, but Forman-Ricci curvature does not take into consideration triangles.  In all cases, it is highly correlated with Augmented Forman-Ricci and Ollivier-Ricci curvature, as seen in Fig. \ref{heatmap_Model} which shows the Spearman correlation of the four model networks with two different sets of parameter values. We can also observe that the correlation between the Bakry-\'Emery-Ricci curvature and other curvatures decreases when the edge density of the networks increases (see Fig. \ref{heatmap_Model} and SI Tables \ref{Modelcorrelation_curvmin_1} and \ref{Modelcorrelation_curvmin_2}). In comparison with Forman-Ricci and Ollivier-Ricci curvature, the correlation between Bakry-\'Emery-Ricci curvature and Augmented Forman-Ricci curvature is stronger. Both Spearman and Pearson correlation for all model networks are listed in the SI Tables \ref{Modelcorrelation_curvmin_1} and \ref{Modelcorrelation_curvmin_2}.

We have also compared the curvature values in the twenty real-world networks. Unlike model networks, Bakry-\'Emery-Ricci curvature exhibits a notable strong correlation with Forman-Ricci, Augmented Forman-Ricci, and Ollivier-Ricci curvature in the majority of real-world networks. The first three columns of Fig. \ref{heatmap_Real} illustrate the Spearman correlation of Bakry-\'Emery-Ricci curvature with Forman-Ricci, Augmented Forman-Ricci, and Ollivier-Ricci curvature for six real-world networks. Compared to Forman-Ricci, Augmented Forman-Ricci, and Ollivier-Ricci curvature display a strong correlation with Bakry-\'Emery-Ricci curvature. Both Spearman and Pearson correlation for all real-world networks are listed in the SI Table \ref{tab:realcorr1_min}.

\subsubsection{Correlation with other vertex measures}

Furthermore, we have compared the curvatures with  other vertex-based measures. These are the  degree, the clustering coefficient, and three centrality measures of the network namely, betweenness centrality, eigenvector centrality \cite{bonacich2007some}, and closeness centrality \cite{okamoto2008ranking}. The degree counts the number of connections a vertex has. The clustering coefficient measure the extent to which the neighbors of a vertex are connected to each other. Formally, the clustering coefficient of a vertex is defined as the ratio of the actual number of triangles to the maximally possible number containing that vertex. Betweenness centrality measures the importance of a vertex within a network for facilitating the flow of information or traffic between other vertices in a network. It quantifies the number of times the vertex lies on the shortest path between other vertices. Eigenvector centrality also measures the importance of a vertex in the networks. It considers not only the number of connections a vertex has but also the importance of those connections. A vertex with high Eigenvector centrality  is connected to other vertices that themselves have high eigenvector centrality. Closeness centrality quantifies the proximity of a vertex to all other vertices within a network. It considers the average shortest path length from a vertex to all other vertices. In other words, it measures how quickly information can spread from a given vertex to others in the network. 

We have observed a high negative correlation between Bakry-\'Emery-Ricci curvature and degree for model networks, specifically, for ER and WS networks. There is a very faint correlation between Bakry-\'Emery-Ricci curvature and degree in BA and HGG models. For all types of model networks, the clustering coefficient does not correlate with the Bakry-\'Emery-Ricci curvature. However, a significant negative correlation is observed between Bakry-\'Emery-Ricci curvature and other vertex centrality measures in the model networks with the exception of the HGG model. Vertices with high centrality measure are known to play an essential role in networks \cite{bringmann2019centrality}. Bakry-\'Emery-Ricci curvature shows a comparatively stronger negative correlation with betweenness centrality in the ER and WS models than in the BA model. However, in three types of model networks, Bakry-\'Emery-Ricci curvature establishes a high negative correlation with eigenvector centrality and closeness centrality. This conveys the fact that vertices with negative curvature have high centrality values. The last five columns of Fig. \ref{heatmap_Model} show the Spearman correlation of Bakry-\'Emery-Ricci curvature with the other vertex-based measures. Also, both the Spearman and Pearson correlation with standard deviation have been provided in SI Tables \ref{Model_Corr_DEGCC}, \ref{Model_Centrality_1}, and \ref{Model_Centrality_2}.

We have also performed a comparison between Bakry-\'Emery-Ricci curvature and other vertex-based measures for real-world networks. Analogous to the model networks, some real-world networks also exhibit a high negative correlation between Bakry-\'Emery-Ricci curvature and degree. Furthermore, the clustering coefficient has no correlation with Bakry-\'Emery-Ricci curvature in most of the real-world networks. The correlation between Bakry-\'Emery-Ricci curvature and centrality measures is moderately to strongly negative. Specifically, the Human protein interaction network shows a strong negative correlation of Bakry-\'Emery-Ricci curvature with eigenvector centrality and closeness centrality compared to betweenness centrality. The last five columns of Fig. \ref{heatmap_Real} show the Spearman correlation of Bakry-\'Emery-Ricci curvature with the other vertex-based measures. Also, both the Spearman and Pearson correlation values for these comparisons have been provided in SI Table \ref{tab:realcorr2}.

\subsection{The relative importance of Bakry-\'Emery-Ricci curvature and topological robustness of networks}

\begin{figure}
    \centering
    \includegraphics{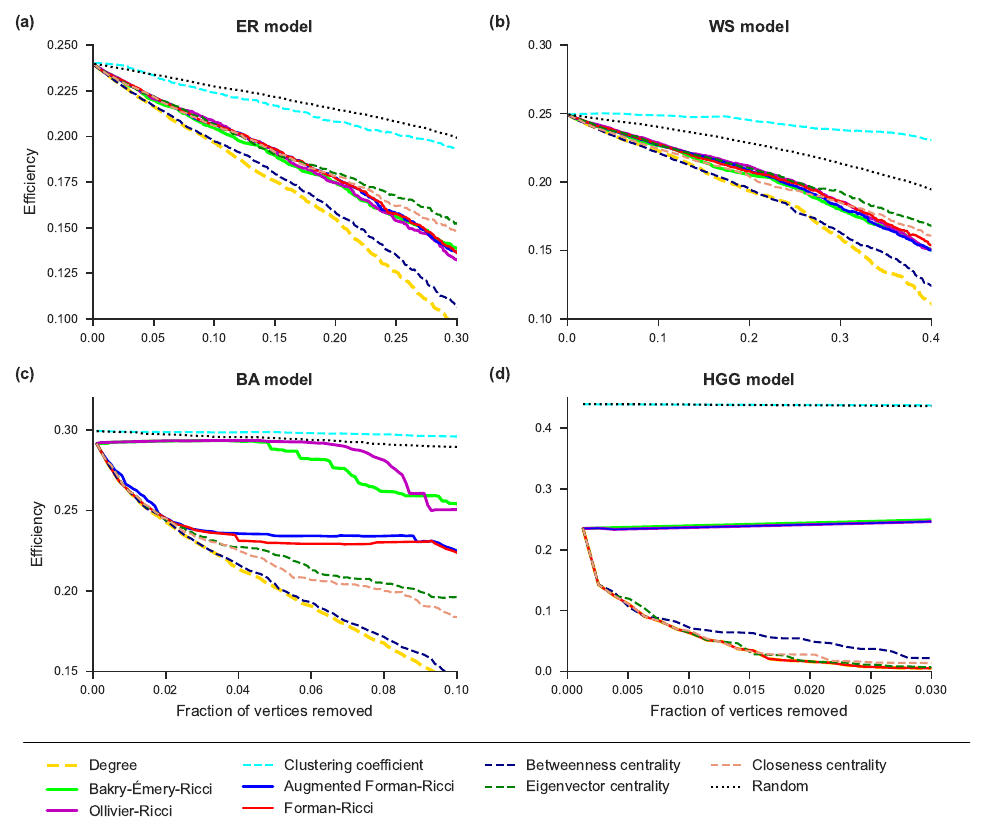}
    \caption{Communication efficiency as function of the fraction of vertices removed in model networks with $n$ = 1000 vertices. (a) ER model with $p$ = 0.005, (b) WS model with $k$ = 6, $p$ = 0.5, (c) BA model with $m$ = 3, and (d) HGG model with $k$ = 3, $\gamma$ = 2, $T$ = 0. Solid lines correspond to different curvature measures, dashed lines correspond to different vertex measures and dotted line corresponds to random vertex removal.}
    \label{Robustness_Model}
\end{figure}

\begin{figure}
    \centering
    \includegraphics{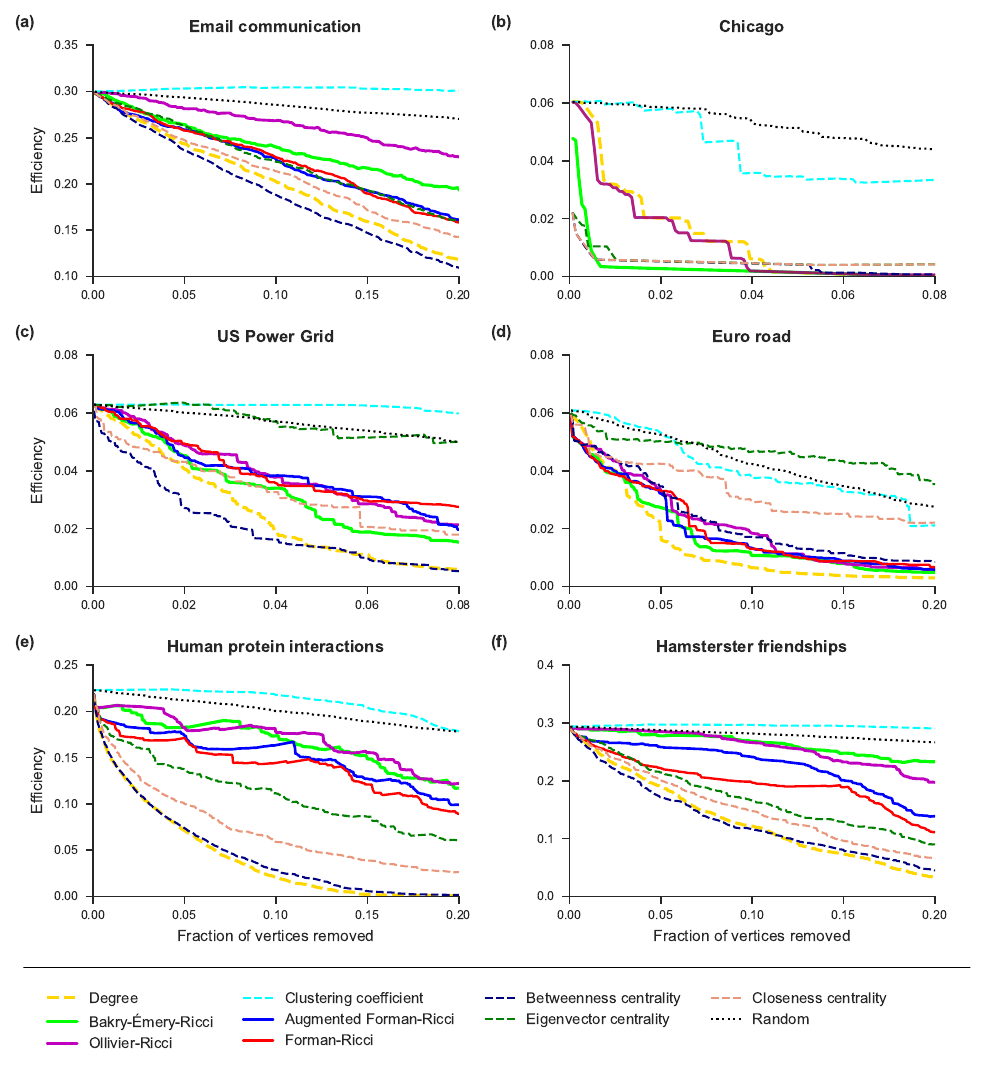}
    \caption{Communication efficiency as function of the fraction of vertices removed in real-world networks (a) Email communication, (b) Chicago road network, (c) US Power Grid, (d) Euro road network, (e) Human protein interactions, and (f) Hamsterster friendships networks. Solid lines correspond to different curvature measures, dashed lines correspond to different vertex measures and dotted line corresponds to random vertex removal.}
    \label{Robustness_Real}
\end{figure}

Communication efficiency is a global network measure to evaluate the impact of vertices removal on the overall connectivity of networks. The communication efficiency of a graph $G$ with $n$ number of vertices is defined as,
\begin{equation}
 E = \frac{1}{n(n-1)}\sum_{i \neq j \in V(G)}{\frac{1}{d_{ij}}}
\end{equation}
where the shortest path between the vertex $i$ and $j$ in the graph $G$ is denoted by $d_{ij}$ and $V(G)$ is the vertex set of the graph $G$. 

We have explored how the communication efficiency of a network changes while removing its vertices. For this, we have considered nine different vertex measures and random vertex removal. We have removed the vertices  one by one and calculated communication efficiency in each time step. Communication efficiency quantifies the sensitivity of the network to a small perturbation in that network. In the case of Bakry-\'Emery-Ricci, Forman-Ricci, Augmented Forman-Ricci, and Ollivier-Ricci curvature we have removed the vertices  according to increasing  curvature values, without reevaluating the curvature for the smaller network. For degree, clustering coefficient, betweenness centrality, eigenvector centrality, and closeness centrality we have removed the vertices depending on the decreasing order of the respective values. For random vertex removal, vertices have been eliminated one by one in a random order. This procedure has been performed ten times for a particular network and the mean values of communication efficiency at each time step are considered.

For model networks, after the removal of vertices with the highest degree, the network shows slightly faster disintegration compared to the other measures. The removal of vertices by increasing  Bakry-\'Emery-Ricci curvature curvature values leads to a faster disintegration of the network compared to random removal and removal depending on decreasing clustering coefficient. In comparison with other curvature measures, Bakry-\'Emery-Ricci curvature exhibits a similar fast disintegration of the networks in ER and WS models. Moreover, the networks are more sensitive after the vertex removal by  decreasing  betweenness centrality than other curvature measures. Since betweenness centrality measures the flow of information through the vertex, the vertices with high values  are important to maintain communication throughout the networks. Further, Eigenvector centrality and closeness centrality exhibit a relatively consistent degradation of the model networks. Fig. \ref{Robustness_Model} demonstrates the importance of vertices in the network depending on their different measures in the model networks.

We have also explored the importance of vertices with highly negative curvature values in real-world networks. Fig. \ref{Robustness_Real} illustrates the communication efficiency for six different real-world networks. In the six networks, removal of vertices with high negative Bakry-\'Emery-Ricci curvature values are more sensitive to the network in comparison to random removal of vertices and vertices with high clustering coefficient. In particular, for the Chicago road network removing vertices based increasing  Bakry-\'Emery-Ricci curvature values leads to a faster disintegration in comparison to removing vertices based on the other measures. Furthermore, vertices with high degree and betweenness centrality play an important role in maintaining the connectivity of the networks.

\subsection{Comparison with scalar curvatures}

\begin{figure}
    \centering
    \includegraphics{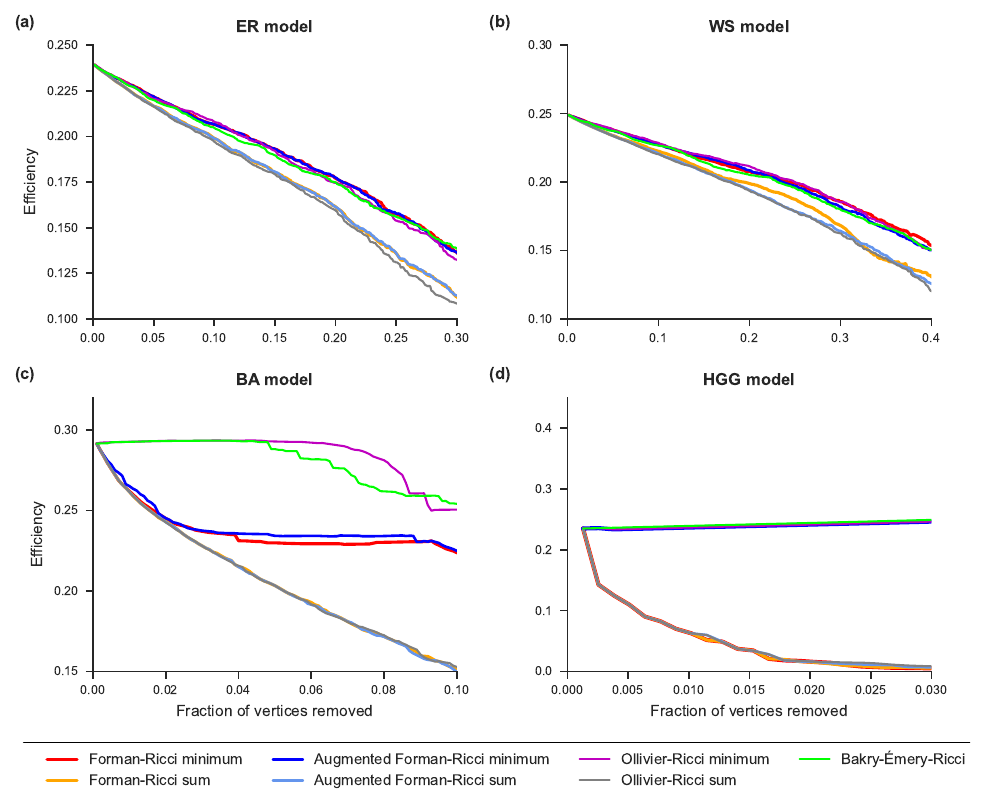}
    \caption{Communication efficiency as function of the fraction of vertices removed in model networks for Forman-Ricci, Augmented Forman-Ricci, Ollivier-Ricci, and Bakry-\'Emery-Ricci curvature. (a) ER model with $p$ = 0.005, (b) WS model with $k$ = 6, $p$ = 0.5, (c) BA model with $m$ = 3, and (d) HGG model with $k$ = 3, $\gamma$ = 2, $T$ = 0. Forman-Ricci minimum, Augmented Forman-Ricci minimum, and Ollivier-Ricci minimum indicate that vertices have been removed depending on the increasing order of $vertex_{min}$ values of Forman-Ricci, Augmented Forman-Ricci, and Ollivier-Ricci curvatures respectively. Forman-Ricci sum, Augmented Forman-Ricci sum, and Ollivier-Ricci sum indicate that vertices have been removed depending on the increasing order of $vertex_{sum}$ values of Forman-Ricci, Augmented Forman-Ricci, and Ollivier-Ricci curvatures respectively. }
    \label{Robustness_Model_curv}
\end{figure}

\begin{figure}
    \centering
    \includegraphics{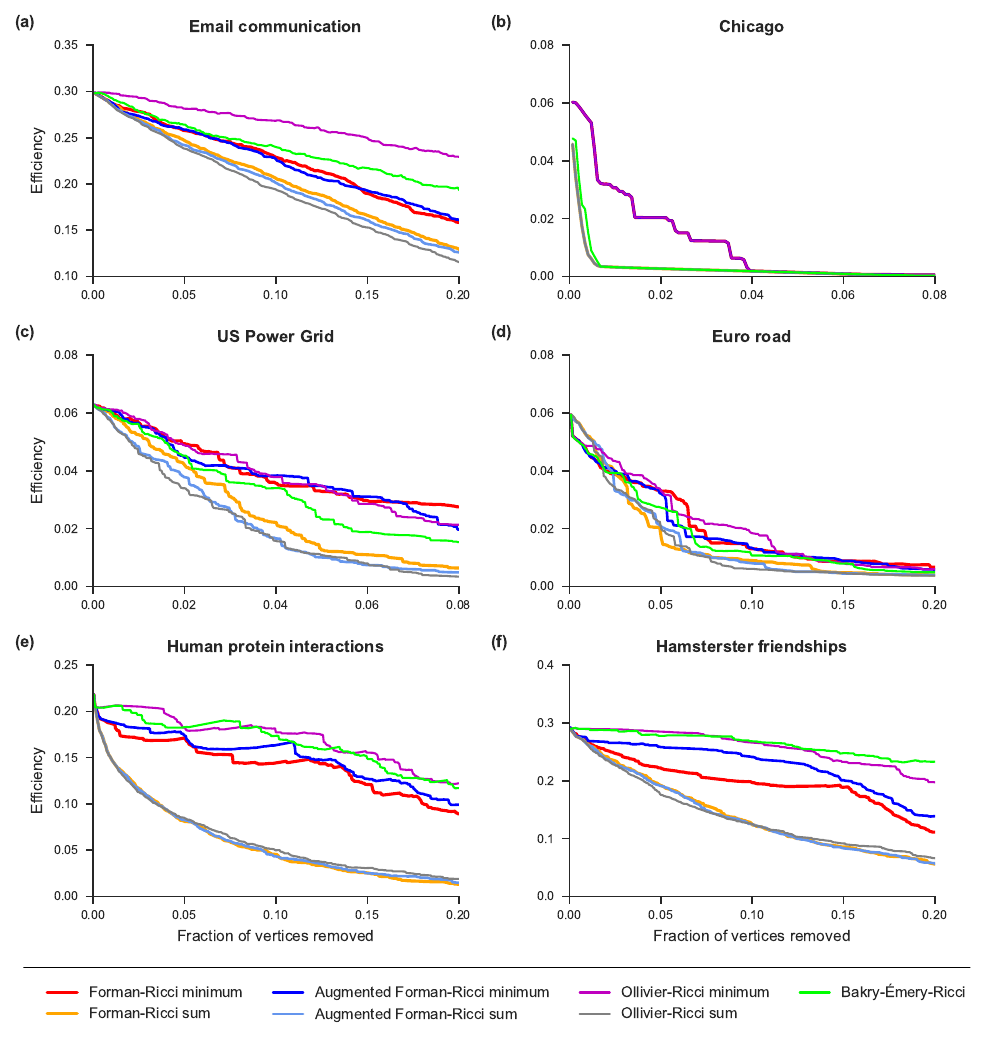}
    \caption{Communication efficiency as function of the fraction of vertices removed in real-world networks for Forman-Ricci, Augmented Forman-Ricci, Ollivier-Ricci, and Bakry-\'Emery-Ricci curvature. (a) Email communication, (b) Chicago road network, (c) US Power Grid, (d) Euro road network, (e) Human protein interactions, and (f) Hamsterster friendships networks. Forman-Ricci minimum, Augmented Forman-Ricci minimum, and Ollivier-Ricci minimum indicate that vertices have been removed depending on the increasing order of $vertex_{min}$ values of Forman-Ricci, Augmented Forman-Ricci, and Ollivier-Ricci curvatures respectively. Forman-Ricci sum, Augmented Forman-Ricci sum, and Ollivier-Ricci sum indicate that vertices have been removed depending on the increasing order of $vertex_{sum}$ values of Forman-Ricci, Augmented Forman-Ricci, and Ollivier-Ricci curvatures respectively.}
    \label{Robustness_Real_curv}
\end{figure}

In this section, we compare $vertex_{sum}$ and $vertex_{min}$ depending on the incident edge curvature values. In defining $vertex_{sum}$ and $vertex_{min}$, we attribute $vertex_{sum}$ to the vertex curvature value obtained as the sum of its incident edges, and $vertex_{min}$ to signify the minimum among its incident edges for the vertex. Previously, we have considered $vertex_{sum}$ \cite{samal2018comparative} which is analogous to scalar curvature in Riemannian geometry \cite{sandhu2015graph}. In Riemannian geometry, scalar curvature captures the change in the volume of a geodesic ball in a curved Riemannian manifold compared to the volume of a standard ball in Euclidean space. Bakry-\'Emery-Ricci curvature also exhibits a higher correlation with the scalar curvatures for both model networks and most of the real-world networks (see SI Tables \ref{Modelcorrelation_curvsum_1}, \ref{Modelcorrelation_curvsum_2}, and \ref{tab:realcorr_sum}). However, between $vertex_{sum}$ and $vertex_{min}$, Bakry-\'Emery-Ricci curvature shows a higher correlation with $vertex_{min}$. Additionally, we have performed the robustness of the networks for Bakry-\'Emery-Ricci curvature and $vertex_{min}$ and $vertex_{sum}$. Figs. \ref{Robustness_Model_curv} and \ref{Robustness_Real_curv} illustrate the importance of vertices depending on their two different Forman-Ricci, Augmented Forman-Ricci, and Ollivier-Ricci vertex curvature measures and Bakry-\'Emery-Ricci curvature in model and real-world networks respectively. For all the cases, $vertex_{sum}$ shows faster disintegration of networks compared to both $vertex_{min}$ and Bakry-\'Emery-Ricci curvature. However, in the majority of networks, the communication efficiency for Bakry-\'Emery-Ricci curvature is approximately similar to $vertex_{min}$ compared to $vertex_{sum}$.

\section{Discussion and conclusions} \label{Discusssion}

We have investigated the properties of Bakry-\'Emery-Ricci curvature, a discrete notion of Ricci curvature, on several undirected and unweighted graph models and real-world networks. We have found that the distribution of Bakry-\'Emery-Ricci curvature for the scale-free and random hyperbolic graphs is broader compared to random and small-world networks. Furthermore, the majority of vertices have a negative curvature value for both model and real-world networks. The Bakry-\'Emery-Ricci curvature is defined on the vertices of a network, whereas the Forman-Ricci, Augmented Forman-Ricci, and Ollivier-Ricci curvatures are defined on the edges of the network. To evaluate Bakry-\'Emery-Ricci curvature of a network with $n$ vertices, a 3D tensor with dimensions $n\times n\times n$ is constructed to make use of fast matrix operations. On the other hand, to obtain the Ollivier-Ricci curvature of the network, one needs to solve a linear programming problem associated with the optimal mass transport for each edge. For a particular network, the computational time for Bakry-\'Emery-Ricci curvature is shorter than that required for Ollivier-Ricci curvature. Furthermore, we have compared the Bakry-\'Emery-Ricci curvature with different discrete notions of Ricci curvature as well as with other vertex measures such as degree, clustering coefficient, and centrality measures. In \cite{pouryahya2017comparing}, it has been shown that the three different notions of discrete Ricci curvature on networks are consistent in cancer networks. Similarly, we have found that the Bakry-\'Emery-Ricci curvature is highly correlated with other discrete notions of Ricci curvature. Nevertheless, when the density of the networks increases, the correlation tends to decrease. Bakry-\'Emery-Ricci curvature is also significantly correlated with the centrality measures for most of the model and real-world networks. Bakry-\'Emery-Ricci curvature shows a particularly strong correlation with Augmented Forman-Ricci curvature. In short, Bakry-\'Emery-Ricci curvature is a useful tool to investigate network properties, and it is simpler to compute than Ollivier-Ricci curvature but somewhat more difficult to compute than Augmented Forman-Ricci curvature. In most networks, Bakry-\'Emery-Ricci curvature is highly correlated with Ollivier-Ricci and Augmented Forman-Ricci, with the exception of HGG model wherein no significant correlation is observed between Bakry-\'Emery-Ricci curvature and other vertex measures. One possible explanation could be that HGG model contains many triangles but there are few vertices with high-degree and most have low-degree. While the values of centrality measures for low-degree vertices is typically low, the Bakry-\'Emery-Ricci curvature becomes very negative for low-degree vertices as they are typically connected with a high-degree node. This is in contrast to the fact that usually, very negative curvature comes with high centrality.

While comparing the robustness, we have found that in ER and WS models, sequential removal of vertices by increasing Bakry-\'Emery-Ricci curvature leads to a similarly fast disintegration as other curvature measures. Additionally, for all the networks it shows faster disintegration compared to clustering coefficient, and random vertex removal. This indicates that Bakry-\'Emery-Ricci curvature may identify vertices that are important to maintain the connectivity of the networks. For Forman-Ricci, Augmented Forman-Ricci, and Ollivier-Ricci curvature we have considered the minimum of the incident edges of one vertex as a vertex measure and compared the robustness with scalar curvature which is analogous to considering the sum of the incident edges as a vertex measure. Thus, vertices with highly negative scalar curvature are more important for the network compared to both Bakry-\'Emery-Ricci curvature and those with a very negative minimum of the incident edges. Since scalar curvature captures the volume growth of small balls, it seems natural that it can identify important vertices of the networks. To analyze this further, one can consider directed and/or weighted networks. Additionally, Bakry-\'Emery-Ricci curvature may be considered to find communities within a network structure. For future investigation of Bakry-\'Emery-Ricci curvature it seems reasonable to focus on high edge density networks. Otherwise, the Augmented Forman-Ricci curvature is easier to compute and expected to give  results that are qualitatively similar to Bakry-\'Emery-Ricci curvature.

\section*{COMPETING INTERESTS}
The authors declare that they have no conflicts of interest.

\section*{AUTHOR CONTRIBUTIONS}
Designed the research: M.M., A.S., F.M., J.J.; Performed the research: M.M., A.S., F.M., J.J.; Performed the computations: M.M.; Wrote the paper: M.M., A.S., F.M., J.J.

\section*{ACKNOWLEDGMENTS}
J.J. acknowledges support from the German-Israeli Foundation, Grant No. I-1514-304.6/2019 ``Discrete curvatures for networks: Comparison and Applications''. A.S. acknowledges support from the Max Planck Society Germany through the award of a Max Planck Partner Group.


\bibliography{refs}

\begin{thebibliography}{10}

\bibitem{networkscience2016}
A.-L. Barabsi.
\newblock {\em Network Science}.
\newblock Cambridge University Press, USA, 2016.

\bibitem{rual2005towards}
J.~Rual, K.~Venkatesan, T.~Hao, T.~Hirozane-Kishikawa, A.~Dricot, N.~Li, G.~F. Berriz, F.~D. Gibbons, M.~Dreze, N.~Ayivi-Guedehoussou, N.~Klitgord, C.~Simon, M.~Boxem, S.~Milstein, J.~Rosenberg, D.~S. Goldberg, L.~V. Zhang, S.~L. Wong, G.~Franklin, S.~Li, J.~S. Albala, J.~Lim, C.~Fraughton, E.~Llamosas, S.~Cevik, C.~Bex, P.~Lamesch, R.~S. Sikorski, J.~Vandenhaute, H.~Y. Zoghbi, A.~Smolyar, S.~Bosak, R.~Sequerra, L.~Doucette-Stamm, M.~E. Cusick, D.~E. Hill, F.~P. Roth, and M.~Vidal.
\newblock Towards a proteome-scale map of the human protein--protein interaction network.
\newblock {\em Nature}, 437(7062):1173--1178, 2005.

\bibitem{eash1979equilibrium}
R.~W. Eash, K.~S. Chon, Y.~J. Lee, and D.~E. Boyce.
\newblock Equilibrium traffic assignment on an aggregated highway network for sketch planning.
\newblock {\em Transportation Research}, 13:243--257, 1979.

\bibitem{vsubelj2011robust}
L.~{\v{S}}ubelj and M.~Bajec.
\newblock Robust network community detection using balanced propagation.
\newblock {\em The European Physical Journal B}, 81(3):353--362, 2011.

\bibitem{leskovec2007graph}
J.~Leskovec, J.~Kleinberg, and C.~Faloutsos.
\newblock Graph {E}volution: {D}ensification and {S}hrinking {D}iameters.
\newblock {\em ACM Trans. Knowl. Discov. Data}, 1(1):2--es, 2007.

\bibitem{sreejith2016forman}
R.~P. Sreejith, K.~Mohanraj, J.~Jost, E.~Saucan, and A.~Samal.
\newblock Forman curvature for complex networks.
\newblock {\em Journal of Statistical Mechanics: Theory and Experiment}, 2016(6):063206, 2016.

\bibitem{samal2018comparative}
A.~Samal, R.~P. Sreejith, J.~Gu, S.~Liu, E.~Saucan, and J.~Jost.
\newblock Comparative analysis of two discretizations of {R}icci curvature for complex networks.
\newblock {\em Scientific Reports}, 8(1):8650, 2018.

\bibitem{cushing2020bakry}
D.~Cushing, S.~Liu, and N.~Peyerimhoff.
\newblock Bakry-\'{E}mery {C}urvature {F}unctions on {G}raphs.
\newblock {\em Canadian Journal of Mathematics}, 72(1):89--143, 2020.

\bibitem{elumalai2022graph}
P.~Elumalai, Y.~Yadav, N.~Williams, E.~Saucan, J.~Jost, and A.~Samal.
\newblock Graph {R}icci curvatures reveal atypical functional connectivity in autism spectrum disorder.
\newblock {\em Scientific Reports}, 12(1):8295, 2022.

\bibitem{yadav2023discrete}
Y.~Yadav, P.~Elumalai, N.~Williams, J.~Jost, and A.~Samal.
\newblock Discrete {R}icci curvatures capture age-related changes in human brain functional connectivity networks.
\newblock {\em Frontiers in Aging Neuroscience}, 15:1120846, 2023.

\bibitem{fesser2023augmentations}
L.~Fesser, S.~S.~H. Iv{\'a}{\~n}ez, K.~Devriendt, M.~Weber, and R.~Lambiotte.
\newblock Augmentations of {F}orman's {R}icci {C}urvature and their {A}pplications in {C}ommunity {D}etection.
\newblock {\em arXiv preprint arXiv:2306.06474}, 2023.

\bibitem{jost2008riemannian}
J.~Jost.
\newblock {\em Riemannian {G}eometry and {G}eometric {A}nalysis}, volume 7th Edn.
\newblock Springer Berlin Heidelberg, 2017.

\bibitem{chow2003combinatorial}
B.~Chow and F.~Luo.
\newblock Combinatorial {R}icci {F}lows on {S}urfaces.
\newblock {\em Journal of Differential Geometry}, 63(1):97--129, 2003.

\bibitem{lott2009ricci}
J.~Lott and C.~Villani.
\newblock Ricci curvature for metric-measure spaces via optimal transport.
\newblock {\em Annals of Mathematics}, 169(3):903--991, 2009.

\bibitem{stone1976combinatorial}
D.~A. Stone.
\newblock A combinatorial analogue of a theorem of {M}yers.
\newblock {\em Illinois Journal of Mathematics}, 20(1):12--21, 1976.

\bibitem{morgan2005manifolds}
F.~Morgan.
\newblock Manifolds with {D}ensity.
\newblock {\em Notices of the Amer. Math. Soc.}, 52(8):853--858, 2005.

\bibitem{bonciocat2009mass}
A.~I. Bonciocat and K.~T. Sturm.
\newblock Mass transportation and rough curvature bounds for discrete spaces.
\newblock {\em Journal of Functional Analysis}, 256(9):2944--2966, 2009.

\bibitem{jost1997nonpositive}
J.~Jost.
\newblock {\em {N}onpositive {C}urvature: {G}eometric and {A}nalytic {A}spects}.
\newblock Springer Science \& Business Media, 1997.

\bibitem{joharinad2019topology}
P.~Joharinad and J.~Jost.
\newblock Topology and curvature of metric spaces.
\newblock {\em Advances in Mathematics}, 356:106813, 2019.

\bibitem{saucan2019discrete}
E.~Saucan, R.~P. Sreejith, R.~P. Vivek-Ananth, J.~Jost, and A.~Samal.
\newblock Discrete {R}icci curvatures for directed networks.
\newblock {\em Chaos, Solitons \& Fractals}, 118:347--360, 2019.

\bibitem{gallot2004riemannian}
S.~Gallot, D.~Hulin, and J.~Lafontaine.
\newblock {\em Riemannian {G}eometry}, volume 3rd Edn.
\newblock Springer Berlin Heidelberg, 2004.

\bibitem{burago2022course}
D.~Burago, Y.~Burago, and S.~Ivanov.
\newblock {\em A {C}ourse in {M}etric {G}eometry}, volume~33.
\newblock American Mathematical Society, 2022.

\bibitem{sher2001handbook}
R.~J. Daverman and R.~B. Sher.
\newblock {\em Handbook of {G}eometric {T}opology}.
\newblock Elsevier, 2002.

\bibitem{lua2005survey}
E.~K. Lua, J.~Crowcroft, M.~Pias, R.~Sharma, and S.~Lim.
\newblock A survey and comparison of peer-to-peer overlay network schemes.
\newblock {\em IEEE Communications Surveys \& Tutorials}, 7(2):72--93, 2005.

\bibitem{saucan2005curvature}
E.~Saucan and E.~Appleboim.
\newblock Curvature {B}ased {C}lustering for {DNA} {M}icroarray {D}ata {A}nalysis.
\newblock In {\em Pattern Recognition and Image Analysis}, volume 3523, pages 405--412. Springer Berlin Heidelberg, 2005.

\bibitem{eidi2020edge}
M.~Eidi, A.~Farzam, W.~Leal, A.~Samal, and J.~Jost.
\newblock Edge-based analysis of networks: curvatures of graphs and hypergraphs.
\newblock {\em Theory in Biosciences}, 139(4):337--348, 2020.

\bibitem{bauer2017geometric}
F.~Bauer, B.~Hua, J.~Jost, S.~Liu, and G.~Wang.
\newblock {\em The {G}eometric {M}eaning of {C}urvature: {L}ocal and {N}onlocal {A}spects of {R}icci {C}urvature}, pages 1--62.
\newblock Springer International Publishing, 2017.

\bibitem{bauer2012ollivier}
F.~Bauer, J.~Jost, and S.~Liu.
\newblock Ollivier-{R}icci curvature and the spectrum of the normalized graph {L}aplace operator.
\newblock {\em Mathematical Research Letters}, 19(06):1185--1205, 2012.

\bibitem{reilly1977applications}
R.~C. Reilly.
\newblock Applications of the {H}essian operator in a {R}iemannian manifold.
\newblock {\em Indiana University Mathematics Department}, 26(3):459--472, 1977.

\bibitem{pouryahya2016bakry}
M.~Pouryahya, R.~Elkin, R.~Sandhu, S.~Tannenbaum, T.~Georgiou, and A.~Tannenbaum.
\newblock Bakry-\'{E}mery {R}icci {C}urvature on {W}eighted {G}raphs with {A}pplications to {B}iological {N}etworks.
\newblock In {\em Int. Symp. on Math. Theory of Net. and Sys}, volume~22, page~52, 2016.

\bibitem{pouryahya2017comparing}
M.~Pouryahya, J.~Mathews, and A.~Tannenbaum.
\newblock {C}omparing {T}hree {N}otions of {D}iscrete {R}icci {C}urvature on {B}iological {N}etworks.
\newblock {\em arXiv preprint arXiv:1712.02943}, 2017.

\bibitem{liu2018bakry}
S.~Liu, F.~M{\"u}nch, and N.~Peyerimhoff.
\newblock Bakry-\'{E}mery curvature and diameter bounds on graphs.
\newblock {\em Calculus of Variations and Partial Differential Equations}, 57(2):67, 2018.

\bibitem{cushing2022bakry}
D.~Cushing, S.~Kamtue, S.~Liu, F.~M{\"u}nch, N.~Peyerimhoff, and H.~B. Snodgrass.
\newblock Bakry-\'{E}mery curvature sharpness and curvature flow in finite weighted graphs. {I}. {T}heory.
\newblock {\em arXiv preprint arXiv:2204.10064}, 2022.

\bibitem{cushing2023bakry}
D.~Cushing, S.~Kamtue, S.~Liu, F.~M{\"u}nch, N.~Peyerimhoff, and B.~Snodgrass.
\newblock Bakry-\'{E}mery {C}urvature {S}harpness and {C}urvature {F}low in {F}inite {W}eighted {G}raphs: {I}mplementation.
\newblock {\em Axioms}, 12(6):577, 2023.

\bibitem{bakry1985diffusions}
D.~Bakry and M.~{\'E}mery.
\newblock Diffusions hypercontractives.
\newblock In {\em S{\'e}minaire de Probabilit{\'e}s XIX 1983/84}, pages 177--206. Springer Berlin Heidelberg, 1985.

\bibitem{schmuckenschlager1999curvature}
M.~Schmuckenschl{\"a}ger.
\newblock Curvature of nonlocal {M}arkov generators.
\newblock {\em Convex Geometric Analysis (Berkeley, CA, 1996)}, 34:189--197, 1999.

\bibitem{hua2017stochastic}
B.~Hua and Y.~Lin.
\newblock Stochastic completeness for graphs with curvature dimension conditions.
\newblock {\em Advances in Mathematics}, 306:279--302, 2017.

\bibitem{jost2014ollivier}
J.~Jost and S.~Liu.
\newblock Ollivier’s {R}icci {C}urvature, {L}ocal {C}lustering and {C}urvature-{D}imension {I}nequalities on {G}raphs.
\newblock {\em Discrete \& Computational Geometry}, 51(2):300--322, 2014.

\bibitem{lin2010ricci}
Y.~Lin and S.~Yau.
\newblock Ricci curvature and eigenvalue estimate on locally finite graphs.
\newblock {\em Math. Res. Lett.}, 17(2):343--356, 2010.

\bibitem{bauer2017curvature}
F.~Bauer, F.~Chung, Y.~Lin, and Y.~Liu.
\newblock {CURVATURE ASPECTS OF GRAPHS}.
\newblock {\em Proceedings of the American Mathematical Society}, 145(5):pp. 2033--2042, 2017.

\bibitem{liu2019distance}
S.~Liu, F.~M{\"u}nch, N.~Peyerimhoff, and C.~Rose.
\newblock Distance {B}ounds for {G}raphs with {S}ome {N}egative {B}akry-\'{E}mery {C}urvature.
\newblock {\em Analysis and Geometry in Metric Spaces}, 7(1):1--14, 2019.

\bibitem{ambrosio2015bakry}
L.~Ambrosio, N.~Gigli, and G.~Savar{\'e}.
\newblock {Bakry-\'{E}mery curvature-dimension condition and {R}iemannian {R}icci curvature bounds}.
\newblock {\em The Annals of Probability}, 43(1):339--404, 2015.

\bibitem{bauer2015li}
F.~Bauer, P.~Horn, Y.~Lin, G.~Lippner, D.~Mangoubi, and S.~Yau.
\newblock Li-{Y}au inequality on graphs.
\newblock {\em Journal of Differential Geometry}, 99(3):359--405, 2015.

\bibitem{horn2019volume}
P.~Horn, Y.~Lin, S.~Liu, and S.~Yau.
\newblock Volume doubling, {P}oincar{\'e} inequality and {G}aussian heat kernel estimate for non-negatively curved graphs.
\newblock {\em Journal f{\"u}r die reine und angewandte Mathematik (Crelles Journal)}, 2019(757):89--130, 2019.

\bibitem{munch2019li}
F.~M{\"u}nch.
\newblock Li-{Y}au inequality under ${CD}(0, n)$ on graphs.
\newblock {\em arXiv preprint arXiv:1909.10242}, 2019.

\bibitem{lin2015equivalent}
Y.~Lin and S.~Liu.
\newblock Equivalent {P}roperties of {CD} {I}nequality on {G}raph.
\newblock {\em arXiv preprint arXiv:1512.02677}, 2015.

\bibitem{keller2018gradient}
M.~Keller and F.~M{\"u}nch.
\newblock Gradient estimates, {B}akry-\'{E}mery {R}icci curvature and ellipticity for unbounded graph {L}aplacians.
\newblock {\em arXiv preprint arXiv:1807.10181}, 2018.

\bibitem{gong2017equivalent}
C.~Gong and Y.~Lin.
\newblock Equivalent properties for {CD} inequalities on graphs with unbounded {L}aplacians.
\newblock {\em Chinese Annals of Mathematics, Series B}, 38(5):1059--1070, 2017.

\bibitem{kempton2022homology}
M.~Kempton, F.~M{\"u}nch, and S.~Yau.
\newblock A homology vanishing theorem for graphs with positive curvature.
\newblock {\em Communications in Analysis and Geometry}, 29(6):1449--1473, 2022.

\bibitem{munch2020spectrally}
F.~M{\"u}nch and C.~Rose.
\newblock Spectrally positive {B}akry-{{\'E}}mery {R}icci curvature on graphs.
\newblock {\em Journal de Math{\'e}matiques Pures et Appliqu{\'e}es}, 143:334--344, 2020.

\bibitem{whitehead1949combinatorial}
J.~H.~C. Whitehead.
\newblock {C}ombinatorial homotopy. {II}.
\newblock {\em Bulletin of the American Mathematical Society}, 55(5):453--496, 1949.

\bibitem{forman2003bochner}
R.~Forman.
\newblock Bochner's {M}ethod for {C}ell {C}omplexes and {C}ombinatorial {R}icci {C}urvature.
\newblock {\em Discrete \& Computational Geometry}, 29(3):323--374, 2003.

\bibitem{jost2021characterizations}
J.~Jost and F.~M{\"u}nch.
\newblock Characterizations of {F}orman curvature.
\newblock {\em arXiv preprint arXiv:2110.04554}, 2021.

\bibitem{saucan2018discrete}
E.~Saucan, A.~Samal, M.~Weber, and J.~Jost.
\newblock Discrete {C}urvatures and {N}etwork {A}nalysis.
\newblock {\em MATCH}, 80(3):605--622, 2018.

\bibitem{sreejith2017systematic}
R.~P. Sreejith, J.~Jost, E.~Saucan, and A.~Samal.
\newblock Systematic evaluation of a new combinatorial curvature for complex networks.
\newblock {\em Chaos, Solitons \& Fractals}, 101:50--67, 2017.

\bibitem{ollivier2007ricci}
Y.~Ollivier.
\newblock Ricci curvature of metric spaces.
\newblock {\em Comptes Rendus Mathematique}, 345(11):643--646, 2007.

\bibitem{ollivier2009ricci}
Y.~Ollivier.
\newblock Ricci curvature of {M}arkov chains on metric spaces.
\newblock {\em Journal of Functional Analysis}, 256(3):810--864, 2009.

\bibitem{ollivier2010survey}
Y.~Ollivier.
\newblock A survey of {R}icci curvature for metric spaces and {M}arkov chains.
\newblock In {\em Probabilistic Approach to Geometry}, pages 343--381. Adv. Stud. Pure Math., 2010.

\bibitem{vaserstein1969markov}
L.~N. Vaserstein.
\newblock Markov {P}rocesses over {D}enumerable {P}roducts of {S}paces, {D}escribing {L}arge {S}ystems of {A}utomata.
\newblock {\em Probl. Peredachi Inf.}, 5(3):64--72, 1969.

\bibitem{munch2019ollivier}
F.~M{\"u}nch and R.~K. Wojciechowski.
\newblock Ollivier {R}icci curvature for general graph {L}aplacians: {H}eat equation, {L}aplacian comparison, non-explosion and diameter bounds.
\newblock {\em Advances in Mathematics}, 356:106759, 2019.

\bibitem{engel2004large}
A.~Engel, R.~Monasson, and A.~K. Hartmann.
\newblock On {L}arge {D}eviation {P}roperties of {E}rd{\"o}s-{R}{\'e}nyi {R}andom {G}raphs.
\newblock {\em Journal of Statistical Physics}, 117(3):387--426, 2004.

\bibitem{watts1998collective}
D.~J. Watts and S.~H. Strogatz.
\newblock Collective dynamics of `small-world’ networks.
\newblock {\em Nature}, 393(6684):440--442, 1998.

\bibitem{barabasi1999emergence}
A.~L. Barab{\'a}si and R.~Albert.
\newblock Emergence of {S}caling in {R}andom {N}etworks.
\newblock {\em Science}, 286(5439):509--512, 1999.

\bibitem{krioukov2010hyperbolic}
D.~Krioukov, F.~Papadopoulos, M.~Kitsak, A.~Vahdat, and M.~Bogu{\~n}{\'a}.
\newblock Hyperbolic geometry of complex networks.
\newblock {\em Phys. Rev. E}, 82(3):036106, 2010.

\bibitem{knuth2006art}
D.~E. Knuth.
\newblock {\em The Art of Computer Programming, Volume 4, Fascicle 4: Generating All Trees--History of Combinatorial Generation}, volume~4.
\newblock Addison-Wesley Professional, 2006.

\bibitem{joshi2005reactome}
G.~Joshi-Tope, M.~Gillespie, I.~Vastrik, P.~D'Eustachio, E.~Schmidt, B.~de~Bono, B.~Jassal, G.R. Gopinath, G.~R. Wu, L.~Matthews, S.~Lewis, E.~Birney, and L.~Stein.
\newblock Reactome: a knowledgebase of biological pathways.
\newblock {\em Nucleic Acids Research}, 33(suppl\_1):D428--D432, 2005.

\bibitem{beuming2005pdzbase}
T.~Beuming, L.~Skrabanek, M.~Y. Niv, P.~Mukherjee, and H.~Weinstein.
\newblock {PDZB}ase: a protein–protein interaction database for {PDZ}-domains.
\newblock {\em Bioinformatics}, 21(6):827--828, 2005.

\bibitem{jeong2001lethality}
H.~Jeong, S.~P. Mason, A.~L. Barab{\'a}si, and Z.~N. Oltvai.
\newblock Lethality and centrality in protein networks.
\newblock {\em Nature}, 411(6833):41--42, 2001.

\bibitem{leskovec2012learning}
J.~McAuley and J.~Leskovec.
\newblock Learning to {D}iscover {S}ocial {C}ircles in {E}go {N}etworks.
\newblock {\em Advances in neural information processing systems}, pages 539--547, 2012.

\bibitem{gleiser2003community}
P.~M. Gleiser and L.~Danon.
\newblock {COMMUNITY STRUCTURE IN JAZZ}.
\newblock {\em Advances in Complex Systems}, 06(04):565--573, 2003.

\bibitem{zachary1977information}
W.~W. Zachary.
\newblock An {I}nformation {F}low {M}odel for {C}onflict and {F}ission in {S}mall {G}roups.
\newblock {\em Journal of Anthropological Research}, 33(4):452--473, 1977.

\bibitem{guimera2003self}
R.~Guimer\`a, L.~Danon, A.~D\'{\i}az-Guilera, F.~Giralt, and A.~Arenas.
\newblock Self-similar community structure in a network of human interactions.
\newblock {\em Phys. Rev. E}, 68(6):065103, 2003.

\bibitem{newman2006finding}
M.~E.~J. Newman.
\newblock Finding community structure in networks using the eigenvectors of matrices.
\newblock {\em Phys. Rev. E}, 74(3):036104, 2006.

\bibitem{boguna2004models}
M.~Bogu\~n{\'a}, R.~Pastor-Satorras, A.~D\'{\i}az-Guilera, and A.~Arenas.
\newblock Models of social networks based on social distance attachment.
\newblock {\em Phys. Rev. E}, 70(5):056122, 2004.

\bibitem{girvan2002community}
M.~Girvan and M.~E.~J. Newman.
\newblock Community structure in social and biological networks.
\newblock {\em Proceedings of the National Academy of Sciences}, 99(12):7821--7826, 2002.

\bibitem{lusseau2003bottlenose}
D.~Lusseau, K.~Schneider, O.~J. Boisseau, P.~Haase, E.~Slooten, and S.~M. Dawson.
\newblock The bottlenose dolphin community of {D}oubtful {S}ound features a large proportion of long-lasting associations.
\newblock {\em Behavioral Ecology and Sociobiology}, 54(4):396--405, 2003.

\bibitem{sundaresan2007network}
S.~R. Sundaresan, I.~R. Fischhoff, J.~Dushoff, and D.~I. Rubenstein.
\newblock Network {M}etrics {R}eveal {D}ifferences in {S}ocial {O}rganization between {T}wo {F}ission-{F}usion {S}pecies, {G}revy's {Z}ebra and {O}nager.
\newblock {\em Oecologia}, 151(1):140--149, 2007.

\bibitem{kunegis2013konect}
J.~Kunegis.
\newblock {KONECT}: {T}he {K}oblenz {N}etwork {C}ollection.
\newblock In {\em Proceedings of the 22nd International Conference on World Wide Web}, pages 1343--1350. Association for Computing Machinery, 2013.

\bibitem{bonacich2007some}
P.~Bonacich.
\newblock Some unique properties of eigenvector centrality.
\newblock {\em Social networks}, 29(4):555--564, 2007.

\bibitem{okamoto2008ranking}
K.~Okamoto, W.~Chen, and X.~Li.
\newblock Ranking of {C}loseness {C}entrality for {L}arge-{S}cale {S}ocial {N}etworks.
\newblock In {\em Frontiers in Algorithmics}, pages 186--195. Springer Berlin Heidelberg, 2008.

\bibitem{bringmann2019centrality}
L.~F. Bringmann, T.~Elmer, S.~Epskamp, R.~W. Krause, D.~Schoch, M.~Wichers, J.~T.~W. Wigman, and E.~Snippe.
\newblock What do centrality measures measure in psychological networks?
\newblock {\em J. Abnorm. Psychol.}, 128(8):892--903, 2019.

\bibitem{sandhu2015graph}
R.~Sandhu, T.~Georgiou, E.~Reznik, L.~Zhu, I.~Kolesov, Y.~Senbabaoglu, and A.~Tannenbaum.
\newblock Graph {C}urvature for {D}ifferentiating {C}ancer {N}etworks.
\newblock {\em Scientific Reports}, 5(1):12323, 2015.

\end{thebibliography}


\onecolumngrid
\setcounter{table}{0}
\renewcommand{\thetable}{S\arabic{table}}
\setcounter{figure}{0}
\renewcommand{\thefigure}{S\arabic{figure}}
\renewcommand{\thesection}{\arabic{section}}


\begin{table}
\centering
\caption {Number of vertices, number of edges, fraction of vertices in the largest connected component (LCC), average degree, edge density and average shortest path length for the model networks considered in this work.}
\begin{tabular}{l >{\cent}p{1.8cm}@{\hspace{0.5cm}} >{\cent}p{1.8cm}@{\hspace{0.5cm}} >{\cent}p{1.8cm}@{\hspace{0.5cm}} >{\cent}p{1.4cm}@{\hspace{0.5cm}} >{\cent}p{1.2cm}@{\hspace{0.5cm}} >{\cent}p{1.9cm}}
\toprule
\toprule
\multirow{3}{*}{\centering \textbf{Model Networks}} & \textbf{Number of Vertices} & \textbf{Number of Edges} & \textbf{Fraction of Vertices in LCC} & \textbf{Average degree} & \textbf{Edge density} & \textbf{Average shortest path length} \\
\midrule
\textbf{ER ($n$ = 1000, $p$ = 0.003)} & 1000 & 1503 & 0.94166 & 3.00616 & 0.00301 & 6.22844 \\
\textbf{ER ($n$ = 1000, $p$ = 0.005)} & 1000 & 2506 & 0.99305 & 5.01254 & 0.00502 & 4.45706 \\
\textbf{ER ($n$ = 1000, $p$ = 0.007)} & 1000 & 3504 & 0.99911 & 7.007 & 0.00701 & 3.76462 \\
\textbf{ER ($n$ = 1000, $p$ = 0.01)} & 1000 & 4992 & 0.99994 & 9.98436 & 0.00999 & 3.25817 \\
\textbf{ER ($n$ = 2000, $p$ = 0.0015)} & 2000 & 3004 & 0.9415 & 3.0039 & 0.0015 & 6.87073 \\
\textbf{ER ($n$ = 2000, $p$ = 0.0025)} & 2000 & 5006 & 0.99314 & 5.00595 & 0.0025 & 4.89227 \\
\textbf{ER ($n$ = 2000, $p$ = 0.0035)} & 2000 & 7002 & 0.99913 & 7.00191 & 0.0035 & 4.12603 \\
\bottomrule
\end{tabular}
\label{Model_details}
\end{table}

\begin{table}[h]
\centering
\caption{Number of vertices, number of edges, fraction of vertices in the largest connected component (LCC), average degree, edge density, and average shortest path length for the real-world networks considered in this work.}
\begin{tabular}{l >{\cent}p{1.8cm}@{\hspace{0.5cm}} >{\cent}p{1.8cm}@{\hspace{0.5cm}} >{\cent}p{1.8cm}@{\hspace{0.5cm}} >{\cent}p{1.4cm}@{\hspace{0.5cm}} >{\cent}p{1.2cm}@{\hspace{0.5cm}} >{\cent}p{1.9cm}}
\toprule
\toprule
\multirow{3}{*}{\centering \textbf{Real-world Networks}} & \textbf{Number of Vertices} & \textbf{Number of Edges} & \textbf{Fraction of Vertices in LCC} & \textbf{Average degree} & \textbf{Edge density} & \textbf{Average shortest path length} \\
\midrule
\textbf{Chicago road} & 1467 & 1298 & 0.56101 & 1.7696 & 0.00121 & 5.91731 \\
\textbf{US Power Grid} & 4941 & 6594 & 1 & 2.6691 & 0.00054 & 18.98919 \\
\textbf{Euro road } & 1174 & 1417 & 0.88501 & 2.41397 & 0.00206 & 18.39515 \\
\textbf{Contiguous US States} & 49 & 107 & 1 & 4.36735 & 0.09099 & 4.16327 \\
\textbf{Human protein interactions} & 2783 & 6007 & 1 & 4.31692 & 0.00155 & 4.8398 \\
\textbf{Reactome} & 6327 & 147547 & 0.94405 & 46.64043 & 0.00737 & 4.21368 \\
\textbf{PDZ-domain} & 212 & 242 & 0.75943 & 2.28302 & 0.01082 & 5.32609 \\
\bottomrule
\end{tabular}
\label{Real_details}
\end{table}

\begin{sidewaystable}
\centering
\caption{Spearman and Pearson correlation along with standard deviation of Bakry-\'Emery-Ricci curvature with Forman-Ricci, Augmented Forman-Ricci, and Ollivier-Ricci curvature of vertices considering the minimum of the incident edges as a vertex measure in the ER and WS model networks generated with different parameter values. The correlations are given up to 5 decimal places.}
\begin{tabular}{p{5.3cm}@{\hspace{0.3cm}}  >{\cent}p{2.8cm} >{\cent}p{2.8cm}  >{\cent}p{2.8cm} >{\cent}p{2.8cm}  >{\cent}p{2.8cm} >{\cent}p{2.8cm}}
\toprule
\toprule

\multirow{2}{*}{\textbf{Model Networks}} & \multicolumn{2}{c}{\textbf{Forman-Ricci}} & \multicolumn{2}{c}{\textbf{Augmented Forman-Ricci}} & \multicolumn{2}{c}{\textbf{Ollivier-Ricci}} \\
 \cmidrule(lr){2-3}
\cmidrule(lr){4-5}
\cmidrule(lr){6-7}
 & \multicolumn{1}{c}{\textbf{Spearman}} & \multicolumn{1}{c}{\textbf{Pearson}} & \multicolumn{1}{c}{\textbf{Spearman}} & \multicolumn{1}{c}{\textbf{Pearson}} & \multicolumn{1}{c}{\textbf{Spearman}} & \multicolumn{1}{c}{\textbf{Pearson}} \\
 \midrule
\textbf{ER ($n$ = 1000, $p$ = 0.003)} &  0.97964 $\pm$ 0.0027 & 0.97626 $\pm$ 0.0025 & 0.98152 $\pm$ 0.0024 & 0.97845 $\pm$ 0.002 & 0.97874 $\pm$ 0.0021 & 0.97652 $\pm$ 0.0018 \\
\textbf{ER ($n$ = 1000, $p$ = 0.005)} & 0.98208 $\pm$ 0.0023 & 0.98093 $\pm$ 0.0025 & 0.9859 $\pm$ 0.0017 & 0.98481 $\pm$ 0.0018 & 0.96871 $\pm$ 0.0027 & 0.97078 $\pm$ 0.0021 \\
\textbf{ER ($n$ = 1000, $p$ = 0.007)} & 0.96944 $\pm$ 0.0032 & 0.97076 $\pm$ 0.003 & 0.976 $\pm$ 0.0022 & 0.97699 $\pm$ 0.0022 & 0.93159 $\pm$ 0.0056 & 0.93912 $\pm$ 0.0046 \\
\textbf{ER ($n$ = 1000, $p$ = 0.01)} & 0.9348 $\pm$ 0.0051 & 0.93903 $\pm$ 0.0047 & 0.94649 $\pm$ 0.0047 & 0.94995 $\pm$ 0.0037 & 0.84104 $\pm$ 0.0126 & 0.85639 $\pm$ 0.01 \\
\textbf{ER ($n$ = 2000, $p$ = 0.0015)} & 0.98244 $\pm$ 0.0016 & 0.97928 $\pm$ 0.0017 & 0.98338 $\pm$ 0.0014 & 0.98039 $\pm$ 0.0013 & 0.98205 $\pm$ 0.0012 & 0.97948 $\pm$ 0.0011 \\
\textbf{ER ($n$ = 2000, $p$ = 0.0025)} & 0.98825 $\pm$ 0.0013 & 0.98734 $\pm$ 0.0011 & 0.99009 $\pm$ 0.0009 & 0.98922 $\pm$ 0.0008 & 0.98092 $\pm$ 0.0012 & 0.98143 $\pm$ 0.0011 \\
\bottomrule
\end{tabular}

\label{Modelcorrelation_curvmin_1}
\end{sidewaystable}

\begin{sidewaystable}
\centering
\caption{Spearman and Pearson correlation along with standard deviation of Bakry-\'Emery-Ricci curvature with Forman-Ricci, Augmented Forman-Ricci, and Ollivier-Ricci curvature of vertices considering the minimum of the incident edges as a vertex measure in the BA and HGG model networks generated with different parameter values. The correlations are given up to 5 decimal places.}
\begin{tabular}{p{4cm}@{\hspace{0.5cm}}  >{\cent}p{2.9cm} >{\cent}p{2.9cm}  >{\cent}p{2.9cm} >{\cent}p{2.9cm}  >{\cent}p{2.9cm} >{\cent}p{2.9cm}}
\toprule
\toprule

\multirow{2}{*}{\textbf{Model Networks}} & \multicolumn{2}{c}{\textbf{Forman-Ricci}} & \multicolumn{2}{c}{\textbf{Augmented Forman-Ricci}} & \multicolumn{2}{c}{\textbf{Ollivier-Ricci}} \\
 \cmidrule(lr){2-3}
\cmidrule(lr){4-5}
\cmidrule(lr){6-7}
 & \multicolumn{1}{c}{\textbf{Spearman}} & \multicolumn{1}{c}{\textbf{Pearson}} & \multicolumn{1}{c}{\textbf{Spearman}} & \multicolumn{1}{c}{\textbf{Pearson}} & \multicolumn{1}{c}{\textbf{Spearman}} & \multicolumn{1}{c}{\textbf{Pearson}} \\
 \midrule
 \textbf{BA ($n$ = 1000, $m$ = 2)} & 0.99368 $\pm$ 0.0013 & 0.98883 $\pm$ 0.0017 & 0.99514 $\pm$ 0.001 & 0.99346 $\pm$ 0.001 & 0.99567 $\pm$ 0.0007 & 0.99676 $\pm$ 0.001 \\
\textbf{BA ($n$ = 1000, $m$ = 3)} & 0.98453 $\pm$ 0.0045 & 0.96267 $\pm$ 0.0039 & 0.98819 $\pm$ 0.0035 & 0.97593 $\pm$ 0.0028 & 0.99244 $\pm$ 0.0017 & 0.99073 $\pm$ 0.002 \\
\textbf{BA ($n$ = 1000, $m$ = 4)} & 0.9599 $\pm$ 0.011 & 0.92396 $\pm$ 0.0065 & 0.96761 $\pm$ 0.0093 & 0.94751 $\pm$ 0.0047 & 0.98419 $\pm$ 0.0042 & 0.97913 $\pm$ 0.003 \\
\textbf{BA ($n$ = 1000, $m$ = 5)} & 0.9153 $\pm$ 0.0174 & 0.87301 $\pm$ 0.0075 & 0.92977 $\pm$ 0.015 & 0.90805 $\pm$ 0.0055 & 0.97073 $\pm$ 0.0059 & 0.96107 $\pm$ 0.0039 \\
\textbf{BA ($n$ = 2000, $m$ = 2)} & 0.99664 $\pm$ 0.0006 & 0.99443 $\pm$ 0.001 & 0.99728 $\pm$ 0.0004 & 0.99656 $\pm$ 0.0006 & 0.99723 $\pm$ 0.0004 & 0.99847 $\pm$ 0.0005 \\
\textbf{BA ($n$ = 2000, $m$ = 3)} & 0.99428 $\pm$ 0.0012 & 0.98058 $\pm$ 0.0021 & 0.9955 $\pm$ 0.001 & 0.98688 $\pm$ 0.0015 & 0.99632 $\pm$ 0.0006 & 0.99568 $\pm$ 0.0009 \\
\bottomrule
\end{tabular}
\label{Modelcorrelation_curvmin_2}
\end{sidewaystable}

 \begin{table}
\centering
\caption{Comparison of Bakry-\'Emery-Ricci curvature with Forman-Ricci, Augmented Forman-Ricci, and Ollivier-Ricci curvature of vertices considering the minimum of the incident edges as a vertex measure in real-world networks. Both Spearman and Pearson correlations are given in this table.}
\begin{tabular}{l >{\cent}p{1.7cm} >{\cent}p{1.7cm} >{\cent}p{2.2cm} >{\cent}p{2.2cm} >{\cent}p{1.7cm} >{\cent}p{1.7cm}}
\toprule
\toprule
\multirow{2}{*}{\textbf{Real-world Networks}} & \multicolumn{2}{c}{\textbf{Forman-Ricci}} & \multicolumn{2}{c}{\textbf{Augmented Forman-Ricci}} & \multicolumn{2}{c}{\textbf{Ollivier-Ricci}} \\
\cmidrule(lr){2-3}
\cmidrule(lr){4-5}
\cmidrule(lr){6-7}
 & \multicolumn{1}{c}{\textbf{Spearman}} & \multicolumn{1}{c}{\textbf{Pearson}} & \multicolumn{1}{c}{\textbf{Spearman}} & \multicolumn{1}{c}{\textbf{Pearson}} & \multicolumn{1}{c}{\textbf{Spearman}} & \multicolumn{1}{c}{\textbf{Pearson}} \\
 \midrule
\textbf{Chicago road} & 0.999476 & 0.997689 & 0.999476 & 0.997689 & 0.999476 & 0.997689 \\
\textbf{US Power Grid} & 0.851368 & 0.852679 & 0.927915 & 0.939042 & 0.929788 & 0.94404 \\
\textbf{Euro road } & 0.925207 & 0.914718 & 0.935944 & 0.929412 & 0.941873 & 0.93907 \\
\textbf{Contiguous US States} & 0.568407 & 0.619974 & 0.795131 & 0.826437 & 0.833624 & 0.885076 \\
\textbf{Human protein interactions} & 0.983131 & 0.959555 & 0.990106 & 0.968287 & 0.993754 & 0.981103 \\
\textbf{Reactome} & 0.447641 & 0.475988 & 0.664213 & 0.782154 & 0.677624 & 0.788065 \\
\bottomrule
\end{tabular}
\label{tab:realcorr1_min}
\end{table}

\begin{table}
\centering
\caption{Spearman and Pearson correlation along with standard deviation of Bakry-\'Emery-Ricci curvature with degree and clustering coefficient in the model networks generated with different parameter values. The correlations are given up to 5 decimal places.}
\begin{tabular}{l >{\cent}p{2.9cm} >{\cent}p{2.9cm}  >{\cent}p{2.9cm} >{\cent}p{2.9cm} }
\toprule
\toprule
\multirow{2}{*}{\textbf{Model Networks}} & \multicolumn{2}{c}{\textbf{Degree}} & \multicolumn{2}{c}{\textbf{Clustering coefficient}} \\
 \cmidrule(lr){2-3}
\cmidrule(lr){4-5}
 & \multicolumn{1}{c}{\textbf{Spearman}} & \multicolumn{1}{c}{\textbf{Pearson}} & \multicolumn{1}{c}{\textbf{Spearman}} & \multicolumn{1}{c}{\textbf{Pearson}} \\
 \midrule
\textbf{ER ($n$ = 1000, $p$ = 0.003)} & -0.88472 $\pm$ 0.0118 & -0.87157 $\pm$ 0.0096 & -0.08176 $\pm$ 0.0385 & 2e-05 $\pm$ 0.0492 \\
\textbf{ER ($n$ = 1000, $p$ = 0.005)} & -0.87648 $\pm$ 0.0117 & -0.87487 $\pm$ 0.0099 & -0.1456 $\pm$ 0.0339 & 0.0085 $\pm$ 0.0458 \\
\textbf{ER ($n$ = 1000, $p$ = 0.007)} & -0.87009 $\pm$ 0.0141 & -0.87349 $\pm$ 0.0123 & -0.18142 $\pm$ 0.0308 & 0.01838 $\pm$ 0.0466 \\
\textbf{ER ($n$ = 1000, $p$ = 0.01)} & -0.86319 $\pm$ 0.0167 & -0.86962 $\pm$ 0.0159 & -0.16258 $\pm$ 0.0354 & 0.04124 $\pm$ 0.0414 \\
\textbf{ER ($n$ = 2000, $p$ = 0.0015)} & -0.88484 $\pm$ 0.0086 & -0.87004 $\pm$ 0.0091 & -0.05914 $\pm$ 0.0258 & -0.00144 $\pm$ 0.0349 \\
\textbf{ER ($n$ = 2000, $p$ = 0.0025)} & -0.87638 $\pm$ 0.0087 & -0.87385 $\pm$ 0.0081 & -0.10324 $\pm$ 0.0259 & 0.00772 $\pm$ 0.0335 \\
\bottomrule
\end{tabular}

\label{Model_Corr_DEGCC}
\end{table}

\begin{sidewaystable}
\centering
\caption{Spearman and Pearson correlation along with standard deviation of Bakry-\'Emery-Ricci curvature with betweenness centrality, eigenvector centrality, and closeness centrality in the ER and WS model networks generated with different parameter values. The correlations are given up to 5 decimal places.}
\begin{tabular}{l  >{\cent}p{2.8cm} >{\cent}p{2.8cm}  >{\cent}p{2.8cm} >{\cent}p{2.8cm}  >{\cent}p{2.8cm} >{\cent}p{2.8cm}}
\toprule
\toprule

\multirow{2}{*}{\textbf{Model Networks}} & \multicolumn{2}{c}{\textbf{Betweenness centrality}} & \multicolumn{2}{c}{\textbf{Eigenvector centrality}} & \multicolumn{2}{c}{\textbf{Closeness centrality}} \\
 \cmidrule(lr){2-3}
\cmidrule(lr){4-5}
\cmidrule(lr){6-7}

 & \multicolumn{1}{c}{\textbf{Spearman}} & \multicolumn{1}{c}{\textbf{Pearson}} & \multicolumn{1}{c}{\textbf{Spearman}} & \multicolumn{1}{c}{\textbf{Pearson}} & \multicolumn{1}{c}{\textbf{Spearman}} & \multicolumn{1}{c}{\textbf{Pearson}} \\
\midrule
\textbf{ER ($n$ = 1000, $p$ = 0.003)} & -0.91644 $\pm$ 0.01 & -0.82651 $\pm$ 0.0087 & -0.88813 $\pm$ 0.0174 & -0.7858 $\pm$ 0.0399 & -0.9121 $\pm$ 0.009 & -0.90395 $\pm$ 0.0086 \\
\textbf{ER ($n$ = 1000, $p$ = 0.005)} & -0.90778 $\pm$ 0.0096 & -0.8475 $\pm$ 0.0095 & -0.93226 $\pm$ 0.0056 & -0.89098 $\pm$ 0.0083 & -0.93443 $\pm$ 0.0056 & -0.93899 $\pm$ 0.0049 \\
\textbf{ER ($n$ = 1000, $p$ = 0.007)} & -0.89674 $\pm$ 0.0117 & -0.85632 $\pm$ 0.0122 & -0.91815 $\pm$ 0.0065 & -0.90125 $\pm$ 0.007 & -0.9238 $\pm$ 0.0068 & -0.92696 $\pm$ 0.0071 \\
\textbf{ER ($n$ = 1000, $p$ = 0.01)} & -0.88132 $\pm$ 0.0149 & -0.85898 $\pm$ 0.0155 & -0.89303 $\pm$ 0.01 & -0.89057 $\pm$ 0.0101 & -0.90572 $\pm$ 0.0098 & -0.91107 $\pm$ 0.0096 \\
\textbf{ER ($n$ = 2000, $p$ = 0.0015)} & -0.92488 $\pm$ 0.0066 & -0.82518 $\pm$ 0.0076 & -0.88791 $\pm$ 0.015 & -0.77714 $\pm$ 0.0419 & -0.90925 $\pm$ 0.0063 & -0.89989 $\pm$ 0.0073 \\
\textbf{ER ($n$ = 2000, $p$ = 0.0025)} & -0.91674 $\pm$ 0.0065 & -0.85166 $\pm$ 0.0074 & -0.93763 $\pm$ 0.0038 & -0.89635 $\pm$ 0.0056 & -0.93462 $\pm$ 0.0043 & -0.93715 $\pm$ 0.0042 \\
\bottomrule
\end{tabular}
\label{Model_Centrality_1}
\end{sidewaystable}

\begin{sidewaystable}
\centering
\caption{Spearman and Pearson correlation along with standard deviation of Bakry-\'Emery-Ricci curvature with betweenness centrality, eigenvector centrality, and closeness centrality in the BA and HGG model networks generated with different parameter values. The correlations are given up to 5 decimal places.}
\begin{tabular}{l  >{\cent}p{2.9cm} >{\cent}p{2.9cm}  >{\cent}p{2.9cm} >{\cent}p{2.9cm}  >{\cent}p{2.9cm} >{\cent}p{2.9cm}}
\toprule
\toprule

\multirow{2}{*}{\textbf{Model Networks}} & \multicolumn{2}{c}{\textbf{Betweenness centrality}} & \multicolumn{2}{c}{\textbf{Eigenvector centrality}} & \multicolumn{2}{c}{\textbf{Closeness centrality}} \\
 \cmidrule(lr){2-3}
\cmidrule(lr){4-5}
\cmidrule(lr){6-7} 

 & \multicolumn{1}{c}{\textbf{Spearman}} & \multicolumn{1}{c}{\textbf{Pearson}} & \multicolumn{1}{c}{\textbf{Spearman}} & \multicolumn{1}{c}{\textbf{Pearson}} & \multicolumn{1}{c}{\textbf{Spearman}} & \multicolumn{1}{c}{\textbf{Pearson}} \\
\midrule
\textbf{BA ($n$ = 1000, $m$ = 2)} & -0.64266 $\pm$ 0.0255 & -0.33756 $\pm$ 0.0488 & -0.8799 $\pm$ 0.0219 & -0.70114 $\pm$ 0.0147 & -0.91063 $\pm$ 0.0124 & -0.88168 $\pm$ 0.0136 \\
\textbf{BA ($n$ = 1000, $m$ = 3)} & -0.60665 $\pm$ 0.0225 & -0.27262 $\pm$ 0.0361 & -0.90208 $\pm$ 0.0091 & -0.59333 $\pm$ 0.0145 & -0.91066 $\pm$ 0.0102 & -0.82692 $\pm$ 0.0157 \\
\textbf{BA ($n$ = 1000, $m$ = 4)} & -0.51574 $\pm$ 0.0231 & -0.23416 $\pm$ 0.0307 & -0.8365 $\pm$ 0.019 & -0.4912 $\pm$ 0.0149 & -0.84299 $\pm$ 0.021 & -0.73292 $\pm$ 0.0156 \\
\textbf{BA ($n$ = 1000, $m$ = 5)} & -0.39728 $\pm$ 0.0271 & -0.2072 $\pm$ 0.0223 & -0.73178 $\pm$ 0.029 & -0.40561 $\pm$ 0.0158 & -0.73865 $\pm$ 0.0314 & -0.62268 $\pm$ 0.0154 \\
\textbf{BA ($n$ = 2000, $m$ = 2)} & -0.67711 $\pm$ 0.0165 & -0.31352 $\pm$ 0.0432 & -0.85289 $\pm$ 0.0217 & -0.70549 $\pm$ 0.0117 & -0.89546 $\pm$ 0.0136 & -0.87477 $\pm$ 0.0142 \\
\textbf{BA ($n$ = 2000, $m$ = 3)} & -0.67094 $\pm$ 0.0134 & -0.25989 $\pm$ 0.0309 & -0.91237 $\pm$ 0.0091 & -0.62303 $\pm$ 0.0123 & -0.92433 $\pm$ 0.0058 & -0.85453 $\pm$ 0.0139 \\
\bottomrule
\end{tabular}

\label{Model_Centrality_2}
\end{sidewaystable}

\begin{sidewaystable}
\centering
\caption{Comparison of Bakry-\'Emery-Ricci curvature with degree, clustering coefficient, betweenness centrality, eigenvector centrality, and closeness centrality in real-world networks. Both Spearman and Pearson correlations are given in this table.}
\begin{tabular}{ l  c  c  c  c  c  c  c  c  c  c }
\toprule
\toprule
\multirow{2}{*}{\textbf{Real-world Networks}} & \multicolumn{2}{c}{\textbf{Degree}} & \multicolumn{2}{c}{\textbf{Clustering coefficient}} & \multicolumn{2}{c}{\textbf{Betweenness centrality}} & \multicolumn{2}{c}{\textbf{Eigenvector centrality}} & \multicolumn{2}{c}{\textbf{Closeness centrality}} \\
 \cmidrule(lr){2-3}
\cmidrule(lr){4-5}
\cmidrule(lr){6-7}
\cmidrule(lr){8-9}
\cmidrule(lr){10-11}
 & \multicolumn{1}{c}{\textbf{Spearman}} & \multicolumn{1}{c}{\textbf{Pearson}} & \multicolumn{1}{c}{\textbf{Spearman}} & \multicolumn{1}{c}{\textbf{Pearson}} & \multicolumn{1}{c}{\textbf{Spearman}} & \multicolumn{1}{c}{\textbf{Pearson}} & \multicolumn{1}{c}{\textbf{Spearman}} & \multicolumn{1}{c}{\textbf{Pearson}} & \multicolumn{1}{c}{\textbf{Spearman}} & \multicolumn{1}{c}{\textbf{Pearson}} \\
\midrule
\textbf{Chicago road} & -0.76548 & -0.86995 & 0 & 0 & -0.76496 & -0.36852 & -0.51196 & -0.66767 & -0.52767 & -0.76824 \\
\textbf{US Power Grid} & -0.68808 & -0.69261 & -0.13287 & 0.128992 & -0.70917 & -0.24837 & -0.381 & -0.09805 & -0.34581 & -0.35668 \\
\textbf{Euro road } & -0.80009 & -0.80452 & -0.25842 & -0.0271 & -0.61655 & -0.42917 & -0.49596 & -0.33316 & -0.54642 & -0.52353 \\
\textbf{Contiguous US States} & -0.71747 & -0.74737 & 0.836074 & 0.647432 & -0.82828 & -0.79695 & -0.4341 & -0.5059 & -0.46869 & -0.50879 \\
\textbf{Human protein interactions} & -0.40261 & -0.29637 & -0.17485 & 0.033314 & -0.42917 & -0.28461 & -0.85907 & -0.53355 & -0.87963 & -0.71439 \\
\textbf{Reactome} & 0.093373 & 0.122283 & -0.07243 & -0.1468 & -0.09058 & -0.17825 & -0.41616 & 0.107816 & -0.36015 & -0.30164 \\
\bottomrule
\end{tabular}
\label{tab:realcorr2}
\end{sidewaystable}

\begin{sidewaystable}
\centering
\caption{Spearman and Pearson correlation along with standard deviation of Bakry-\'Emery-Ricci curvature with Forman-Ricci, Augmented Forman-Ricci, and Ollivier-Ricci curvature of vertices considering the sum of the incident edges as a vertex measure in the ER and WS model networks generated with different parameter values. The correlations are given up to 5 decimal places.}
\begin{tabular}{p{5.3cm}@{\hspace{0.3cm}}  >{\cent}p{2.8cm} >{\cent}p{2.8cm}  >{\cent}p{2.8cm} >{\cent}p{2.8cm}  >{\cent}p{2.8cm} >{\cent}p{2.8cm}}
\toprule
\toprule
\multirow{2}{*}{\textbf{Model Networks}} & \multicolumn{2}{c}{\textbf{Forman-Ricci}} & \multicolumn{2}{c}{\textbf{Augmented Forman-Ricci}} & \multicolumn{2}{c}{\textbf{Ollivier-Ricci}} \\
\cmidrule(lr){2-3}
\cmidrule(lr){4-5}
\cmidrule(lr){6-7}
 & \multicolumn{1}{c}{\textbf{Spearman}} & \multicolumn{1}{c}{\textbf{Pearson}} & \multicolumn{1}{c}{\textbf{Spearman}} & \multicolumn{1}{c}{\textbf{Pearson}} & \multicolumn{1}{c}{\textbf{Spearman}} & \multicolumn{1}{c}{\textbf{Pearson}} \\
 \midrule
\textbf{ER ($n$ = 1000, $p$ = 0.003)} & 0.96512 $\pm$ 0.0039 & 0.85102 $\pm$ 0.0091 & 0.96647 $\pm$ 0.0038 & 0.85242 $\pm$ 0.0091 & 0.96651 $\pm$ 0.0036 & 0.85321 $\pm$ 0.0091 \\
\textbf{ER ($n$ = 1000, $p$ = 0.005)} & 0.93097 $\pm$ 0.0069 & 0.86728 $\pm$ 0.0093 & 0.93257 $\pm$ 0.0067 & 0.86885 $\pm$ 0.0092 & 0.93015 $\pm$ 0.0068 & 0.86857 $\pm$ 0.0095 \\
\textbf{ER ($n$ = 1000, $p$ = 0.007)} & 0.90892 $\pm$ 0.01 & 0.87249 $\pm$ 0.0115 & 0.91085 $\pm$ 0.0098 & 0.87437 $\pm$ 0.0115 & 0.90681 $\pm$ 0.0106 & 0.87242 $\pm$ 0.0123 \\
\textbf{ER ($n$ = 1000, $p$ = 0.01)} & 0.88779 $\pm$ 0.0135 & 0.87089 $\pm$ 0.0147 & 0.8902 $\pm$ 0.0135 & 0.87338 $\pm$ 0.0146 & 0.88879 $\pm$ 0.0138 & 0.86985 $\pm$ 0.0159 \\
\textbf{ER ($n$ = 2000, $p$ = 0.0015)} & 0.96633 $\pm$ 0.0029 & 0.84964 $\pm$ 0.0096 & 0.96701 $\pm$ 0.0029 & 0.85024 $\pm$ 0.0094 & 0.96704 $\pm$ 0.0029 & 0.85062 $\pm$ 0.0095 \\
\textbf{ER ($n$ = 2000, $p$ = 0.0025)} & 0.93296 $\pm$ 0.0055 & 0.86835 $\pm$ 0.0079 & 0.93373 $\pm$ 0.0055 & 0.86906 $\pm$ 0.0079 & 0.93226 $\pm$ 0.0055 & 0.86894 $\pm$ 0.008 \\
\bottomrule
\end{tabular}

\label{Modelcorrelation_curvsum_1}
\end{sidewaystable}

\begin{sidewaystable}
\centering
\caption{Spearman and Pearson correlation along with standard deviation of Bakry-\'Emery-Ricci curvature with Forman-Ricci, Augmented Forman-Ricci, and Ollivier-Ricci curvature of vertices considering the sum of the incident edges as a vertex measure in the BA and HGG model networks generated with different parameter values. The correlations are given up to 5 decimal places.}
\begin{tabular}{p{4cm}@{\hspace{0.5cm}}  >{\cent}p{2.9cm} >{\cent}p{2.9cm}  >{\cent}p{2.9cm} >{\cent}p{2.9cm}  >{\cent}p{2.9cm} >{\cent}p{2.9cm}}
\toprule
\toprule

\multirow{2}{*}{\textbf{Model Networks}} & \multicolumn{2}{c}{\textbf{Forman-Ricci}} & \multicolumn{2}{c}{\textbf{Augmented Forman-Ricci}} & \multicolumn{2}{c}{\textbf{Ollivier-Ricci}} \\
 \cmidrule(lr){2-3}
\cmidrule(lr){4-5}
\cmidrule(lr){6-7}
 & \multicolumn{1}{c}{\textbf{Spearman}} & \multicolumn{1}{c}{\textbf{Pearson}} & \multicolumn{1}{c}{\textbf{Spearman}} & \multicolumn{1}{c}{\textbf{Pearson}} & \multicolumn{1}{c}{\textbf{Spearman}} & \multicolumn{1}{c}{\textbf{Pearson}} \\
 \midrule
\textbf{BA ($n$ = 1000, $m$ = 2)} & 0.94304 $\pm$ 0.0064 & 0.32972 $\pm$ 0.0542 & 0.94399 $\pm$ 0.0064 & 0.33007 $\pm$ 0.0546 & 0.94862 $\pm$ 0.0064 & 0.32813 $\pm$ 0.0551 \\
\textbf{BA ($n$ = 1000, $m$ = 3)} & 0.86719 $\pm$ 0.012 & 0.28148 $\pm$ 0.0374 & 0.86806 $\pm$ 0.0118 & 0.28242 $\pm$ 0.0379 & 0.89064 $\pm$ 0.011 & 0.28068 $\pm$ 0.0385 \\
\textbf{BA ($n$ = 1000, $m$ = 4)} & 0.76365 $\pm$ 0.0215 & 0.25507 $\pm$ 0.0305 & 0.76427 $\pm$ 0.0211 & 0.25641 $\pm$ 0.031 & 0.81495 $\pm$ 0.0198 & 0.25823 $\pm$ 0.0317 \\
\textbf{BA ($n$ = 1000, $m$ = 5)} & 0.63736 $\pm$ 0.0319 & 0.23421 $\pm$ 0.0215 & 0.63731 $\pm$ 0.0317 & 0.23543 $\pm$ 0.0217 & 0.71726 $\pm$ 0.0307 & 0.24425 $\pm$ 0.0223 \\
\textbf{BA ($n$ = 2000, $m$ = 2)} & 0.95279 $\pm$ 0.0044 & 0.29643 $\pm$ 0.0496 & 0.95324 $\pm$ 0.0043 & 0.29657 $\pm$ 0.0498 & 0.95532 $\pm$ 0.0043 & 0.29484 $\pm$ 0.0502 \\
\textbf{BA ($n$ = 2000, $m$ = 3)} & 0.89905 $\pm$ 0.0082 & 0.25762 $\pm$ 0.0345 & 0.89939 $\pm$ 0.0082 & 0.25811 $\pm$ 0.0348 & 0.9115 $\pm$ 0.0079 & 0.25529 $\pm$ 0.0354 \\
\bottomrule
\end{tabular}

\label{Modelcorrelation_curvsum_2}
\end{sidewaystable}

\begin{table}
\centering
\caption{Comparison of Bakry-\'Emery-Ricci curvature with Forman-Ricci, Augmented Forman-Ricci, and Ollivier-Ricci curvature of vertices considering the sum of the incident edges as a vertex measure in real-world networks. Both Spearman and Pearson correlations are given in this table.}
\begin{tabular}{l >{\cent}p{1.7cm} >{\cent}p{1.7cm} >{\cent}p{2.2cm} >{\cent}p{2.2cm} >{\cent}p{1.7cm} >{\cent}p{1.7cm}}
\toprule
\toprule
\multirow{2}{*}{\textbf{Real-world Networks}} & \multicolumn{2}{c}{\textbf{Forman-Ricci}} & \multicolumn{2}{c}{\textbf{Augmented Forman-Ricci}} & \multicolumn{2}{c}{\textbf{Ollivier-Ricci}} \\
\cmidrule(lr){2-3}
\cmidrule(lr){4-5}
\cmidrule(lr){6-7}
 & \multicolumn{1}{c}{\textbf{Spearman}} & \multicolumn{1}{c}{\textbf{Pearson}} & \multicolumn{1}{c}{\textbf{Spearman}} & \multicolumn{1}{c}{\textbf{Pearson}} & \multicolumn{1}{c}{\textbf{Spearman}} & \multicolumn{1}{c}{\textbf{Pearson}} \\
 \midrule
\textbf{Chicago road} & 0.956374 & 0.863091 & 0.956374 & 0.863091 & 0.956374 & 0.863091 \\
\textbf{US Power Grid} & 0.843028 & 0.650911 & 0.92355 & 0.719759 & 0.926952 & 0.71008 \\
\textbf{Euro road } & 0.923635 & 0.78261 & 0.932844 & 0.792534 & 0.940625 & 0.798089 \\
\textbf{Contiguous US States} & 0.64181 & 0.708634 & 0.762168 & 0.774334 & 0.74203 & 0.777042 \\
\textbf{Human protein interactions} & 0.892565 & 0.29271 & 0.912513 & 0.292284 & 0.935504 & 0.285448 \\
\textbf{Reactome} & 0.134264 & -0.04148 & 0.645721 & 0.278316 & 0.771247 & 0.313906 \\
\bottomrule
\end{tabular}

\label{tab:realcorr_sum}
\end{table}

\begin{figure}[h]
    \centering
    \includegraphics{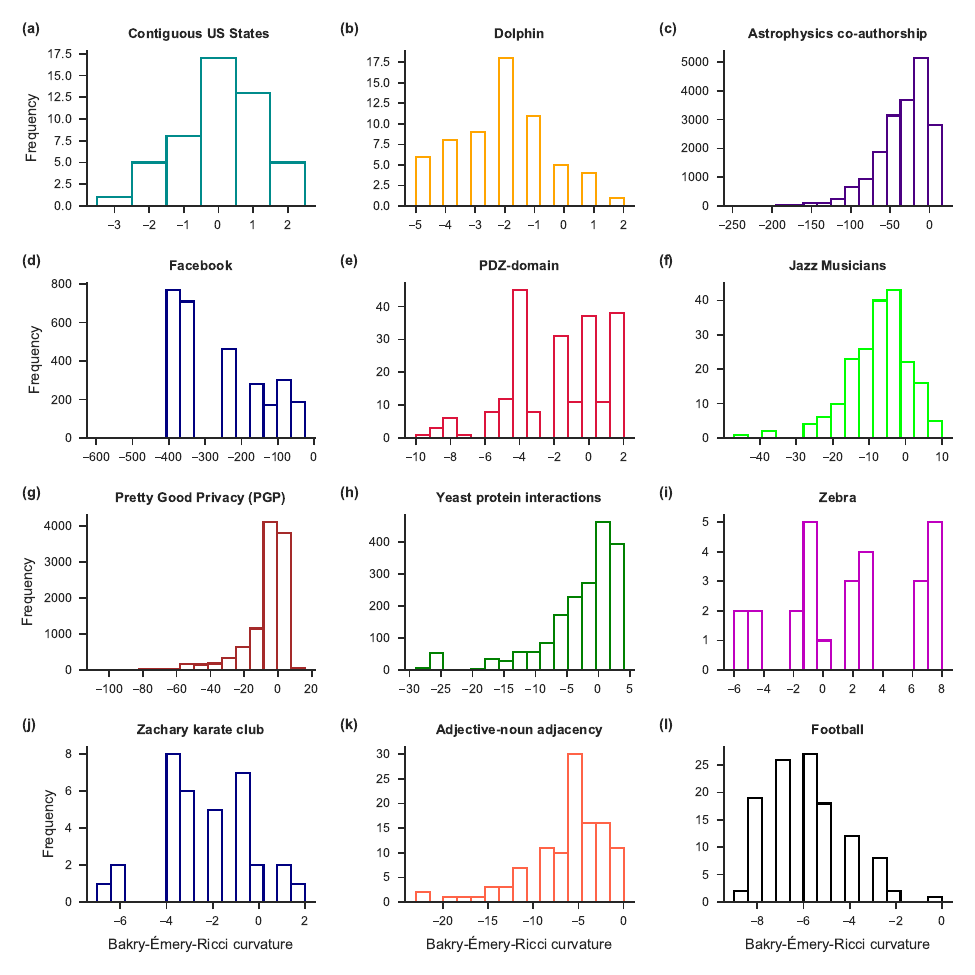}
    \caption{Distribution of Bakry-\'Emery-Ricci curvature for different real-world networks. (a) Contiguous US States network, (b) Dolphin network, (c) Astrophysics co-authorship network, (d) Facebook network, (e) PDZ-domain network, (f) Jazz Musicians network, (g) Pretty Good Privacy (PGP) network, (h) Yeast protein interactions network, (i) Zebra network, (j) Zachary karate club network, (k) Adjective-noun adjacency network, and (l) Football network.}
    \label{Distribution_realothers}
\end{figure}

\end{document}